\documentclass[aps,prd,amsmath,superscriptaddress,nofootinbib,%
               onecolumn,12pt,tightenlines]{revtex4}

\usepackage{graphicx}

\usepackage{ifthen}

\newcommand{\coloronlinestatement}{}

\newcommand{\ie}{\textit{i.e.}}
\newcommand{\eg}{\textit{e.g.}}

\newcommand{\rmd}{\ensuremath{\mathrm{d}}}
\newcommand{\Real}{\mathop{\textrm{Re}}}
\newcommand{\diag}{\mathop{\textrm{diag}}}
\newcommand{\orderof}[1]{\ensuremath{\mathcal{O}(#1)}}

\newcommand{\gae}{%
  \ensuremath{\lower 2pt \hbox{%
    $\, \buildrel {\scriptstyle >}\over {\scriptstyle \sim}\,$}%
    }%
  }
\newcommand{\lae}{%
  \ensuremath{\lower 2pt \hbox{%
    $\, \buildrel {\scriptstyle <}\over {\scriptstyle \sim}\,$}%
    }%
  }

\newcommand{\refeqn}[2][eqn:]{Eqn.~(\ref{#1#2})}

\newcommand{\reftab}[2][tab:]{Table~\ref{#1#2}}

\newcommand{\reffig}[2][fig:]{Figure~\ref{#1#2}}
\newcommand{\Reffig}[2][fig:]{Figure~\ref{#1#2}}
\newcommand{\refsec}[2][sec:]{Section~\ref{#1#2}} 

\newcommand{\ifmulticol}[2]{%
  \ifthenelse{\lengthtest{1.9\columnwidth<\textwidth}}{#1}{#2}%
}

\newcommand{\scfigwidth}{%
  \ifdim1.9\columnwidth<\textwidth%
    1.00\columnwidth\else0.75\columnwidth\fi%
}

\newcommand{\tanb}{\ensuremath{\tan{\beta}}}
\newcommand{\signmu}{\ensuremath{{\rm sign}(\mu)}}

\newcommand{\rmu}{\ensuremath{\mathrm{u}}}
\newcommand{\rms}{\ensuremath{\mathrm{s}}}
\newcommand{\rmc}{\ensuremath{\mathrm{c}}}
\newcommand{\rmb}{\ensuremath{\mathrm{b}}}
\newcommand{\rmt}{\ensuremath{\mathrm{t}}}
\newcommand{\rmpp}{\ensuremath{\mathrm{p}}}
\newcommand{\rmnn}{\ensuremath{\mathrm{n}}}

\newcommand{\mup}{\ensuremath{m_{\rm u}}}
\newcommand{\mumd}{\ensuremath{m_{\rm u} / m_{\rm d}}}
\newcommand{\md}{\ensuremath{m_{\rm d}}}
\newcommand{\ms}{\ensuremath{m_{\rm s}}}
\newcommand{\msmd}{\ensuremath{m_{\rm s} / m_{\rm d}}}
\newcommand{\mc}{\ensuremath{m_{\rm c}}}
\newcommand{\mb}{\ensuremath{m_{\rm b}}}
\newcommand{\mt}{\ensuremath{m_{\rm t}}}
\newcommand{\SigmapiN}{\ensuremath{\Sigma_{\pi\!{\scriptscriptstyle N}}}}
\newcommand{\athree}{\ensuremath{a_{3}^{\rm (p)}}}
\newcommand{\aeight}{\ensuremath{a_{8}^{\rm (p)}}}
\newcommand{\Deltaps}{\ensuremath{\Delta_{\rms}^{(\rmpp)}}}
\newcommand{\sigmaN}{\ensuremath{\sigma_{\chi N}}}
\newcommand{\sigmaNSI}{\ensuremath{\sigma_{\chi N,\rm SI}}}
\newcommand{\sigmapSI}{\ensuremath{\sigma_{\chi\rm p,SI}}}
\newcommand{\sigmanSI}{\ensuremath{\sigma_{\chi\rm n,SI}}}
\newcommand{\sigmaNSD}{\ensuremath{\sigma_{\chi N,\rm SD}}}
\newcommand{\sigmapSD}{\ensuremath{\sigma_{\chi\rm p,SD}}}
\newcommand{\sigmanSD}{\ensuremath{\sigma_{\chi\rm n,SD}}}
\newcommand{\rhoDM}{\ensuremath{\rho_{0}}}
\newcommand{\fTq}[1]{\ensuremath{f_{T_{#1}}}}
\newcommand{\fNTq}[1]{\ensuremath{f_{T_{#1}}^{(N)}}}
\newcommand{\fpTq}[1]{\ensuremath{f_{T_{#1}}^{(\rmpp)}}}
\newcommand{\fnTq}[1]{\ensuremath{f_{T_{#1}}^{(\rmnn)}}}
\newcommand{\Bq}[1]{\ensuremath{B_{#1}}}
\newcommand{\BNq}[1]{\ensuremath{B_{#1}^{(N)}}}
\newcommand{\Bpq}[1]{\ensuremath{B_{#1}^{(\rmpp)}}}
\newcommand{\Bnq}[1]{\ensuremath{B_{#1}^{(\rmnn)}}}

\newcommand{\DeltaNq}[1]{\ensuremath{\Delta_{#1}^{(N)}}}
\newcommand{\Deltapq}[1]{\ensuremath{\Delta_{#1}^{(\rmpp)}}}
\newcommand{\Deltanq}[1]{\ensuremath{\Delta_{#1}^{(\rmnn)}}}

\begin{document}


\preprint{CERN-PH-TH/2008-005}
\preprint{UMN--TH--2631/08}
\preprint{FTPI--MINN--08/02}
\rightline{CERN-PH-TH/2008-005}
\rightline{UMN--TH--2631/08}
\rightline{FTPI--MINN--08/02}


\title{Hadronic Uncertainties in the Elastic Scattering of
       Supersymmetric Dark Matter}

\author{\bf John Ellis}
\email[]{John.Ellis@cern.ch}
\affiliation{
 TH Division, Physics Department,
 CERN,
 1211 Geneva 23, Switzerland}

\author{\bf Keith A. Olive}
\email[]{olive@physics.umn.edu}
\affiliation{
 William I. Fine Theoretical Physics Institute,
 School of Physics and Astronomy,
 University of Minnesota,
 Minneapolis, MN 55455, USA}

\author{\bf Christopher Savage}
\email[]{cmsavage@physics.umn.edu}
\affiliation{
 William I. Fine Theoretical Physics Institute,
 School of Physics and Astronomy,
 University of Minnesota,
 Minneapolis, MN 55455, USA}

\date{\today}


\vskip 1in

\begin{abstract}
\vskip .5in

\centerline{\bf Abstract}

\vskip .2in

We review the uncertainties in the spin-independent and -dependent
elastic scattering cross sections of supersymmetric dark matter
particles on protons and neutrons. We propagate the uncertainties in
quark masses and hadronic matrix elements that are related to the
$\pi$-nucleon $\sigma$ term and the spin content of the nucleon. By
far the largest single uncertainty is that in spin-independent
scattering induced by our ignorance of the $\langle N | {\bar q} q | N
\rangle$ matrix elements linked to the $\pi$-nucleon $\sigma$ term,
which affects the ratio of cross sections on proton and neutron
targets as well as their absolute values. This uncertainty is already
impacting the interpretations of experimental searches for cold dark
matter. {\it We plead for an
experimental campaign to determine better the $\pi$-nucleon $\sigma$
term.}  Uncertainties in the spin content of the proton affect
significantly, but less strongly, the calculation of rates used in
indirect searches.

\end{abstract}

\maketitle

\vfill
\leftline{CERN-PH-TH/2008-005}
\leftline{January 2008}


\section{\label{sec:intro} Introduction}

The most convincing way to confirm the existence and nature of dark
matter particles would be to observe directly their scattering on
nuclei in low-background underground experiments. The sensitivities of
these experiments are currently improving rapidly, and beginning to
cut into the parameter space of plausible supersymmetric
scenarios~\cite{osusy07}.  In order to evaluate accurately the impacts
of these experiments, it is important to understand and minimize the
hadronic uncertainties in the elastic scattering matrix elements for
any given supersymmetric model. The rates for elastic scattering also
control the rates for the capture of dark matter particles by celestial
bodies such the Sun or Earth. There are good prospects for increasing
significantly the sensitivities of experiments looking indirectly for
astrophysical dark matter via the products of their annihilations in
such bodies, adding to the motivations for understanding and reducing
their uncertainties.

Beyond the interpretation of upper limits on dark matter scattering
may lie the interpretation of any eventual detection of a signal and
the task of identifying the nature of the dark matter particle. In
principle, there are four observables that could contribute to such an
analysis, namely the spin-independent and -dependent cross sections on
protons and neutrons, respectively. Part of the strategy for
identifying the nature of any detected dark matter signal would be the
comparison of the measured rates for scattering on different targets,
with the comparison of spin-independent and -dependent scattering
rates playing a particularly important role as a diagnostic
tool~\cite{Savage:2004fn,Bertone:2007xj}. As we see later, currently
there are considerable uncertainties also in such comparisons, related
principally to uncertainties in the hadronic matrix elements of
higher-dimensional effective interactions.

In this paper, we consider only hadronic uncertainties in the elastic
scattering rates.  There are also potentially important uncertainties
related to the supersymmetric model itself, namely how accurately one
can estimate the coefficient of a given higher-dimensional effective
interaction in a given model, and also in the astrophysical density of
dark matter particles. Reducing the model uncertainty would require,
\eg, a complete calculation of radiative corrections to the effective
scattering operator, which lies beyond the scope of this work. As for
the local density of cold dark matter, it is usually taken to be
0.3~GeV/cm$^{3}$, but lower values have occasionally been advocated.

The hadronic uncertainties we consider are listed in
\reftab{params}. They include those in the quark masses,
expressed as $m_{\rm{d,c,b,t}}$ and the ratios $\mumd$ and $\msmd$,
those in the matrix elements $\langle N | {\bar q} q | N \rangle$,
which are related to the change in the nucleon mass
due to non-zero quark masses, denoted by $\sigma_0$, and 
therefore to the $\pi$-nucleon
$\sigma$ term, $\SigmapiN$ as discussed later, and the
axial-current matrix elements $\langle N | {\bar q} \gamma_\mu
\gamma_5 q | N \rangle$, which are related to the quantities
$\Deltaps, \athree$ and $\aeight$, as also discussed later. We find
that the uncertainties in the elastic scattering cross section induced
by the uncertainties in the quark masses, apart from the top quark,
are negligible. However, cross section uncertainties
induced by the uncertainties in the matrix elements $\langle N |
{\bar q} q | N \rangle$ and $\langle N | {\bar q} \gamma_\mu \gamma_5
q | N \rangle$ are very important, as we discuss below. In particular,
the uncertainties induced by our ignorance of $\SigmapiN$ are
particularly important.

\begin{table}
  \addtolength{\tabcolsep}{1em}
  \begin{tabular}{lc@{$\,\pm\,$}cc}
    \hline \hline
    \mumd & 0.553 & 0.043    &~\cite{Leutwyler:1996qg} \\
    \md   & 5     & 2 MeV    &~\cite{Yao:2006px} \\
    \msmd & 18.9  & 0.8      &~\cite{Leutwyler:1996qg} \\
    \mc   & 1.25  & 0.09 GeV &~\cite{Yao:2006px} \\
    \mb   & 4.20  & 0.07 GeV &~\cite{Yao:2006px} \\
    \mt   & 171.4 & 2.1 GeV  &~\cite{Heinson:2006yq} \\
    \hline
    $\sigma_0$ & 36 & 7 MeV &~\cite{Borasoy:1996bx} \\
    \SigmapiN  & 64 & 8 MeV &~\cite{Pavan:2001wz,Ellis:2005mb} \\
    \hline
    \athree  & 1.2695 & 0.0029 &~\cite{Yao:2006px} \\
    \aeight  & 0.585  & 0.025  &~\cite{Goto:1999by,Leader:2002ni} \\
    \Deltaps & -0.09  & 0.03   &~\cite{Alekseev:2007vi} \\
    \hline \hline
  \end{tabular}
  \caption[Parameters]{
    Hadronic parameters used to determine neutralino-nucleon
    scattering cross-sections, with estimates of their experimental
    uncertainties.
    }
  \label{tab:params}
\end{table}

For illustration, we work within the minimal supersymmetric extension
of the Standard Model (MSSM) with conserved $R$ parity, and assume that
the astrophysical cold dark matter is provided by the lightest
neutralino $\chi$~\cite{EHNOS}. We further assume a constrained MSSM
(CMSSM) framework, in which the soft supersymmetry-breaking parameters
$m_{1/2}, m_0$ and $A_0$ are assumed to be universal at the GUT input
scale, and restrict our attention to scenarios with the Higgs mixing
parameter $\mu > 0$ and specific values of the ratio of supersymmetric
Higgs v.e.v.s $\tanb$~\cite{cmssm}.  We illustrate our observations by
studies of some specific CMSSM benchmark scenarios~\cite{bench}, and
also by surveys along strips in the $(m_{1/2}, m_0)$ plane for
$\tan \beta = 10, 50$ along which ${\tilde \tau} - \chi$
coannihilation maintains the relic neutralino density within the range
favoured by WMAP and other experiments~\cite{cmssmwmap}.

We find that the spin-independent cross section may vary by almost an
order of magnitude for 48~MeV $< \SigmapiN < 80$~MeV, the $\pm
2$-$\sigma$ range according to the uncertainties in
\reftab{params}. This uncertainty is already
impacting the interpretations of experimental searches for cold dark
matter. Propagating the $\pm 2$-$\sigma$ uncertainties
in $\Deltaps$, the next most important parameter, we find a variation
by a factor $\sim 2$ in the spin-dependent cross section. Since the
spin-independent cross section may now be on the verge of
detectability in certain models, and the uncertainty in the cross
section is far greater, {\it we appeal for a greater, dedicated effort
to reduce the experimental uncertainty in the $\pi$-nucleon $\sigma$
term $\SigmapiN$}. This quantity is not just an object of curiosity
for those interested in the structure of the nucleon and
non-perturbative strong-interaction effects: it may also be key to
understanding new physics beyond the Standard Model.

\section{\label{sec:MSSM} Supersymmetric Framework}

We briefly review in this Section the theoretical framework we use in
the context of the MSSM; for more comprehensive reviews, see,
\eg,~\cite{Nilles:1983ge,Haber:1984rc}. The neutralino LSP is the
lowest-mass eigenstate combination of the Bino ${\tilde B}$, Wino 
$\tilde W$ and Higgsinos ${\tilde H}_{1,2}$, whose mass matrix $N$ is
diagonalized by a matrix $Z$: $\diag(m_{\chi_{1,..,4}}) = Z^* N Z^{-1}$.
The composition of the lightest neutralino may be written as
\begin{equation} \label{eqn:chi}
  \chi = Z_{\chi 1}\tilde{B} + Z_{\chi 2}\tilde{W} +
         Z_{\chi 3}\tilde{H_{1}} + Z_{\chi 4}\tilde{H_{2}}.
\end{equation}
As already mentioned, we work here in the context of the CMSSM and
assume universality at the supersymmetric GUT scale for the gaugino
masses, $m_{1/2}$,  as well as for the soft scalar masses, $m_0$,
and tri-linear terms, $A_0$.  Our treatment of the sfermion mass
matrices $M$ follows~\cite{Falk:1998xj,Falk:1999mq}.  The sfermion
mass-squared matrix is diagonalized by a matrix $\eta$:
$diag(m^2_1, m^2_2) \equiv \eta M^2 \eta^{-1}$, which can be
parameterized for each flavour $f$ by an angle $\theta_{f}$. We ignore
here all possible CP-violating phases.  The diagonalization matrix can
be written as
\begin{equation} \label{eqn:etamatrix}
  \left( \begin{array}{cc}
    \cos{\theta_{f}} & \sin{\theta_{f}} e^{i \gamma_{f}} \\
    -\sin{\theta_{f}} e^{-i \gamma_{f}} & \cos{\theta_{f}}
  \end{array} \right) 
  \hspace{0.5cm}
  \equiv 
  \hspace{0.5cm}
  \left( \begin{array}{cc}
    \eta_{11} & \eta_{12} \\
    \eta_{21} & \eta_{22}
  \end{array} \right).
\end{equation}
The magnitudes of $\mu$ and the pseudoscalar Higgs mass $m_A$ are
calculated from the electroweak vacuum conditions using
$\mt = 171.4$~GeV, except as noted in \refsec{results}.

The only four-fermi Lagrangian contributions for describing elastic
$\chi$-nucleon scattering obtained from the MSSM Lagrangian which are
not velocity dependent, and hence relevant for relic dark matter
scattering, are~\cite{Falk:1998xj}:
\begin{equation} \label{eqn:lagrangian}
  {\cal L}
    = \alpha_{2i} \bar{\chi} \gamma^\mu \gamma^5 \chi \bar{q_{i}} 
    \gamma_{\mu} \gamma^{5}  q_{i}
+ \alpha_{3i} \bar{\chi} \chi \bar{q_{i}} q_{i}.
\end{equation}
This Lagrangian is to be summed over the quark generations, and the 
subscript $i$ labels up-type quarks ($i=1$) and down-type quarks
($i=2$).   The coefficients are given by:
\begin{eqnarray} \label{eqn:alpha2}
  \alpha_{2i}
    &=& \frac{1}{4(m^{2}_{1i} - m^{2}_{\chi})}
        \left[ \left| X_{i} \right|^{2} + \left| Y_{i} \right|^{2} \right] 
    \ifmulticol{\nonumber \\ &&}{}
        {}+ \frac{1}{4(m^{2}_{2i} - m^{2}_{\chi})}
            \left[ \left| W_{i} \right|^{2}
                   + \left| V_{i} \right|^{2} \right]
        \nonumber \\
    & & {}- \frac{g^{2}}{4 m_{Z}^{2} \cos^{2}{\theta_{W}}}
            \left[ \left| Z_{\chi_{3}} \right|^{2}
                   - \left| Z_{\chi_{4}} \right|^{2} \right]
            \frac{T_{3i}}{2}
\end{eqnarray}
and
\begin{eqnarray} \label{eqn:alpha3}
  \alpha_{3i}
    &=& {}- \frac{1}{2(m^{2}_{1i} - m^{2}_{\chi})}
            \Real \left[ \left( X_{i} \right)
                         \left( Y_{i} \right)^{\ast} \right] 
    \ifmulticol{\nonumber \\ &&}{}
        {}- \frac{1}{2(m^{2}_{2i} - m^{2}_{\chi})}
            \Real \left[ \left( W_{i} \right)
                         \left( V_{i} \right)^{\ast} \right]
        \nonumber \\
    & & {}- \frac{g m_{q_{i}}}{4 m_{W} B_{i}}
    \ifmulticol{\nonumber\\ && \quad \times}{}
        \bigg\{
          \left( \frac{D_{i}^{2}}{m^{2}_{H_{2}}}
                 + \frac{C_{i}^{2}}{m^{2}_{H_{1}}}
          \right)
          \Real \left[ \delta_{2i}
                  \left( g Z_{\chi 2} - g' Z_{\chi 1}\right)
                \right]
    \ifmulticol{\nonumber\\ && \quad\quad\quad}%
               {\nonumber\\ && \qquad\qquad\qquad}
          {}+ D_{i} C_{i} \left( \frac{1}{m^{2}_{H_{2}}}
                                 - \frac{1}{m^{2}_{H_{1}}} \right)
    \ifmulticol{\nonumber\\ && \quad\quad\quad\quad\;\: \times}{}
              \Real \left[ \delta_{1i}
                      \left( g Z_{\chi 2} - g' Z_{\chi 1} \right)
                    \right]
        \bigg\},
\end{eqnarray}
where
\begin{eqnarray} \label{eqn:XYWV}
  X_{i} &\equiv&
    \eta^{\ast}_{11} \frac{g m_{q_{i}}Z_{\chi 5-i}^{\ast}}{2 m_{W} B_{i}}
    - \eta_{12}^{\ast} e_{i} g' Z_{\chi 1}^{\ast}
    \nonumber \\
  Y_{i} &\equiv&
    \eta^{\ast}_{11} \left( \frac{y_{i}}{2} g' Z_{\chi 1}
                            + g T_{3i} Z_{\chi 2} \right)
    + \eta^{\ast}_{12} \frac{g m_{q_{i}} Z_{\chi 5-i}}{2 m_{W} B_{i}}
    \nonumber \\
  W_{i} &\equiv&
    \eta_{21}^{\ast} \frac{g m_{q_{i}}Z_{\chi 5-i}^{\ast}}{2 m_{W} B_{i}}
    - \eta_{22}^{\ast} e_{i} g' Z_{\chi 1}^{\ast}
    \nonumber \\
  V_{i} &\equiv&
    \eta_{21}^{\ast}\left( \frac{y_{i}}{2} g' Z_{\chi 1}
                           + g T_{3i} Z_{\chi 2} \right)
    + \eta_{22}^{\ast} \frac{g m_{q_{i}} Z_{\chi 5-i}}{2 m_{W} B_{i}},
\end{eqnarray}
where $y_i, T_{3i}$ denote hypercharge and isospin, and
\begin{eqnarray} \label{eqn:deltas}
  \delta_{1i} = Z_{\chi 3} (Z_{\chi 4}),
  \qquad
  \delta_{2i} = Z_{\chi 4} (-Z_{\chi 3}),
\end{eqnarray}
\begin{eqnarray} \label{eqn:CD}
  B_{i}             &=&     \sin{\beta} (\cos{\beta}),
  \ifmulticol{\nonumber \\}{\qquad}
  C_{i} \ifmulticol{&=&}{=} \sin{\alpha} (\cos{\alpha}),
  \qquad
  D_{i}              =      \cos{\alpha} (-\sin{\alpha}) 
\end{eqnarray}
for up (down) type quarks.  We denote by $m_{H_2} < m_{H_1}$ the masses
of the two neutral scalar Higgs bosons, and $\alpha$ denotes the
neutral Higgs boson mixing angle.

\section{\label{sec:hadron} Hadronic Matrix Elements}

The elastic cross section for neutralino scattering off a nucleus can be
decomposed into a scalar (spin-independent) part obtained from
the $\alpha_{3i}$ term in \refeqn{lagrangian}, and a spin-dependent
part obtained from the $\alpha_{2i}$ term.  Each of these can be
written in terms of the cross sections for elastic scattering for
scattering off individual nucleons, as we now review and re-evaluate.

\subsection{\label{sec:hadronSI} Spin-Independent Term}

The scalar, or spin-independent (SI), part of the cross section can be
written as~\footnote{This expression is valid in the
         zero-momentum-transfer limit.  For non-zero momentum exchange,
         the expression must include a form factor due to the finite
         size of the nucleus.  See, for example, Ref.~\cite{SmithLewin}.}%
\begin{equation} \label{eqn:sigmaSI}
  \sigma_{\rm SI} = \frac{4 m_{r}^{2}}{\pi}
                    \left[ Z f_{p} + (A-Z) f_{n}  \right]^{2},
\end{equation}
where $m_r$ is the $\chi$-nuclear reduced mass and
\begin{equation} \label{eqn:fN}
  \frac{f_N}{m_N}
    = \sum_{q=\rmu,\rmd,\rms} \fNTq{q} \frac{\alpha_{3q}}{m_{q}}
      + \frac{2}{27} f_{TG}^{(N)}
        \sum_{q=\rmc,\rmb,\rmt} \frac{\alpha_{3q}}{m_q}
\end{equation}
for $N$ = p or n.  The parameters $\fNTq{q}$ are defined by
\begin{equation} \label{eqn:Bq}
  m_N \fNTq{q}
  \equiv \langle N | m_{q} \bar{q} q | N \rangle
  \equiv m_q \BNq{q} ,
\end{equation}
where~\cite{Shifman:1978zn,Vainshtein:1980ea}
\begin{equation} \label{eqn:fTG}
  f_{TG}^{(N)} = 1 - \sum_{q=\rmu,\rmd,\rms} \fNTq{q} .
\end{equation}
We take the ratios of the light quark masses from
\cite{Leutwyler:1996qg}:
\begin{equation} \label{eqn:mqmd}
  \frac{\mup}{\md} = 0.553 \pm 0.043 , \qquad
  \frac{\ms}{\md}  = 18.9 \pm 0.8 .
\end{equation}
We take the other quark masses from~\cite{Yao:2006px}, except for the
top mass, which is taken from the combined CDF and D0 result
\cite{Heinson:2006yq}.  These masses, as well as other experimental
quantities that will arise in the calculation of the hadronic matrix
elements, appear in \reftab{params}.

Following~\cite{Cheng:1988im}, we introduce the quantity:
\begin{equation} \label{eqn:z}
  z \equiv \frac{\Bpq{\rmu} - \Bpq{\rms}}{\Bpq{\rmd} - \Bpq{\rms}}
    = 1.49 ,
\end{equation}
which has an experimental error that is negligible compared with others
discussed below, and the strange scalar density
\begin{equation} \label{eqn:y}
  y \equiv \frac{2 \BNq{\rms}}{\BNq{\rmu} + \BNq{\rmd}}.
\end{equation}
In terms of these, one may write
\begin{equation} \label{eqn:BdBu}
  \frac{\Bpq{\rmd}}{\Bpq{\rmu}}
   = \frac{2 + ((z - 1) \times y)}{2 \times z - ((z - 1) \times y)} \; .
\end{equation}
Proton and neutron scalar matrix elements are related by an interchange
of $B_{\rmu}$ and $B_{\rmd}$, i.e.,
\begin{equation} \label{eqn:Bn}
  \Bnq{\rmu} = \Bpq{\rmd} , \quad
  \Bnq{\rmd} = \Bpq{\rmu} , \quad \text{and} \quad
  \Bnq{\rms} = \Bpq{\rms} .
\end{equation}
The $\pi$-nucleon sigma term, $\SigmapiN$, may be written as
\begin{equation} \label{eqn:SigmapiN}
  \SigmapiN \equiv \frac{1}{2} (\mup + \md)
                   \times \left( \BNq{\rmu} + \BNq{\rmd} \right) \; ,
\end{equation}
and the coefficients $\fTq{q}$ may be written in the forms:
\begin{alignat}{3}
  \label{eqn:fNTu}
  \fTq{\rmu}
    & \ = \ & \frac{\mup \Bq{\rmu}}{m_N}
    & \ = \ & \frac{2 \SigmapiN}{m_N (1+\frac{\md}{\mup})
                                 (1+\frac{\Bq{\rmd}}{\Bq{\rmu}})}
    \; , \\
  \label{eqn:fNTd}
  \fTq{\rmd}
    & \ = \ & \frac{\md \Bq{\rmd}}{m_N}
    & \ = \ & \frac{2 \SigmapiN}{m_N (1+\frac{\mup}{\md})
                                 (1+\frac{\Bq{\rmu}}{\Bq{\rmd}})}
    \; , \\
  \label{eqn:fNTs}
  \fTq{\rms}
    & \ = \ & \frac{\ms \Bq{\rms}}{m_N}
    & \ = \ & \frac{(\frac{\ms}{\md}) \SigmapiN \, y}%
                   {m_N (1+\frac{\mup}{\md})}
    \; ; \quad\quad\quad
\end{alignat}
where we have dropped the $(N)$ superscript from $\fTq{q}$ and
$\Bq{q}$.

The effect of the uncertainties in the $\fTq{q}$ were considered
in~\cite{priorvals,priorvals2} and we were motivated to
reconsider~\cite{Ellis:2005mb} the value of $y$ by recent
re-evaluations of  the $\pi$-nucleon sigma term $\SigmapiN$, which is
related to the strange scalar density in the nucleon by
\begin{equation} \label{eqn:y2}
  y = 1 - \sigma_0/\SigmapiN \; .
\end{equation}
The value for $\sigma_0$ given in \reftab{params} is estimated on the
basis of octet baryon mass differences to be $\sigma_0 = 36 \pm 7$~MeV
\cite{Borasoy:1996bx,Gasser:1990ce,Knecht:1999dp,Sainio:2001bq}.
Recent determinations of $\SigmapiN$ have found the following values
at the Cheng-Dashen point $t = + 2 m_\pi^2$ \cite{Pavan:2001wz}:
\begin{equation} \label{eqn:SigmaCD}
  \Sigma_{CD} \, = \, (88 \pm 15, \, 71 \pm 9, \, 79 \pm 7,
                       \, 85 \pm 5) \;\text{MeV} \; .
\end{equation}
These should be corrected by an amount $- \Delta_R - \Delta_\sigma
\simeq - 15$~MeV to obtain $\SigmapiN$. Assuming for definiteness the
value $\Sigma_{CD} = 79 \pm 7$~MeV, one finds
\begin{equation} \label{eqn:SigmapiNexp}
  \SigmapiN \, = \, (64 \pm 8) \; \text{MeV} \; .
\end{equation}
This is the range generally considered in this paper, though one could
even argue for a larger uncertainty, and we also discuss the
implications if $\SigmapiN = \sigma_0$, \ie, $y = 0$.

\subsection{\label{sec:hadronSD} Spin-Dependent Term}

The spin-dependent (SD) part of the elastic $\chi$-nucleus cross
section can be written as~\footnote{As with the SI cross section, this
        expression applies in the zero momentum transfer limit and
        requires an additional form factor for finite momentum
        transfer.  This form factor may have a small but non-zero
        dependence on $a_{\rmpp}$ and $a_{\rmnn}$.}%
\begin{equation} \label{eqn:sigmaSD}
  \sigma_{\rm SD} = \frac{32}{\pi} G_{F}^{2} m_{r}^{2}
                    \Lambda^{2} J(J + 1) \; ,
\end{equation}
where $m_{r}$ is again the reduced neutralino mass, $J$ is the spin 
of the nucleus,
\begin{equation} \label{eqn:Lambda}
  \Lambda \equiv \frac{1}{J} \left(
                 a_{\rmpp} \langle S_{\rmpp} \rangle
                 + a_{\rmnn} \langle S_{\rmnn} \rangle
                 \right) \; ,
\end{equation}
and
\begin{equation} \label{eqn:aN}
  a_{\rmpp} = \sum_{q} \frac{\alpha_{2q}}{\sqrt{2} G_{f}}
              \Deltapq{q} , \qquad
  a_{\rmnn} = \sum_{i} \frac{\alpha_{2q}}{\sqrt{2} G_{f}}
              \Deltanq{q} \; .
\end{equation}
The factors $\DeltaNq{q}$ parametrize the quark spin content of the
nucleon and are only significant for the light (u,d,s) quarks.
A combination of experimental and theoretical results tightly
constrain the linear combinations~\cite{Yao:2006px}
\begin{equation} \label{eqn:a3}
  \athree \equiv \Deltapq{\rmu} - \Deltapq{\rmd}
          = 1.2695 \pm 0.0029
\end{equation}
and~\cite{Goto:1999by,Leader:2002ni}
\begin{equation} \label{eqn:a8}
  \aeight \equiv \Deltapq{\rmu} + \Deltapq{\rmd} - 2 \Deltapq{\rms}
          = 0.585 \pm 0.025 .
\end{equation}
However, the individual $\DeltaNq{q}$ are relatively poorly
constrained; using the recent COMPASS result~\cite{Alekseev:2007vi},
\begin{eqnarray} \label{eqn:Deltaps}
  \Deltapq{\rms} &=& -0.09 \pm 0.01 \, \text{(stat.)} \,
                         \pm 0.02 \, \text{(syst.)}
  \ifmulticol{\nonumber \\ &\approx&}{\approx}
                     -0.09 \pm 0.03 \; ,
\end{eqnarray}
where we have conservatively combined the statistical and systematic
uncertainties, we may express $\DeltaNq{\rmu,\rmd}$ as follows in terms
of known quantities:
\begin{alignat}{3}
  \label{eqn:Deltapu}
  \Deltapq{\rmu}
    & \ = \ & \frac{1}{2} \left( \aeight + \athree \right) + \Deltaps
    & \ = \ & 0.84 \pm 0.03
    \; , \\
  \label{eqn:Deltapd}
  \Deltapq{\rmd}
    & \ = \ & \frac{1}{2} \left( \aeight - \athree \right) + \Deltaps
    & \ = \ & -0.43 \pm 0.03
    \; .
\end{alignat}
The above two uncertainties and that of $\Deltaps$, however, are
correlated and we shall instead use the independent quantities
$\athree$ and $\aeight$ when appropriate.
These values differ by approximately 2$\sigma$ from those used in 
\cite{priorvals} and we will explore the impact of this change on
the magnitude and ratios of the spin-dependent cross sections.
The proton and neutron scalar matrix elements are related by an
interchange of $\Delta_{\rmu}$ and $\Delta_{\rmd}$, or
\begin{equation} \label{eqn:Deltan}
  \Deltanq{\rmu} = \Deltapq{\rmd} , \quad
  \Deltanq{\rmd} = \Deltapq{\rmu} , \quad \text{and} \quad
  \Deltanq{\rms} = \Deltapq{\rms} .
\end{equation}

\section{\label{sec:results} Scattering cross sections, ratios and
         uncertainties}

Direct and some indirect dark matter detection techniques involve the
scattering of a WIMP off a nucleus.  Direct detection experiments, such
as
CDMS~\cite{Akerib:2005kh,Akerib:2005za},
XENON10~\cite{xenon,xenonSD},
ZEPLIN-II~\cite{Alner:2007ja,Alner:2007xs},
and KIMS~\cite{Lee:2007qn}
aim to detect dark matter via scattering of relic neutralinos off
nuclei inside the detectors~\cite{Goodman:1984dc}.  Indirect detection
experiments such as
Super-Kamiokande~\cite{Desai:2004pq}
and AMANDA/IceCube~\cite{Silvestri:2007zza},
on the other hand, search for high-energy neutrinos produced in WIMP
annihilations at the center of the Sun~\cite{indirectdet:solar}
or Earth~\cite{indirectdet:earth}.  The annihilations are the result
of WIMPs accumulating at the centers of these massive bodies due to
galactic WIMPs scattering off nuclei in these bodies and losing enough
energy to become gravitationally bound. Subsequent scatters then cause
the WIMPs to fall to the cores, where the local density is enhanced.
Other indirect detection methods search for WIMPs that annihilate in
the Galactic Halo or near the Galactic Center, where they produce
neutrinos, positrons, or antiprotons that may be seen in detectors on
the Earth~\cite{indirectdet:galactichalo,indirectdet:galacticcenter}.
However, such annihilations do not involve scattering off nuclei,
and will not be discussed here.

The rate of such scattering events and the relative
sensitivities of different experiments are dependent upon the four
$\chi$-nucleon scattering cross-sections (spin-independent (SI) and
spin-dependent (SD) cross-sections for each of the proton and neutron).
The interpretation of any positive signal in such an experiment
requires an understanding of the precision that such a signal can be
correlated with underlying parameters within a given theoretical
framework (\eg, $m_0$, $m_{1/2}$, and $\tanb$ in the CMSSM).

Direct and indirect detection signals are inherently proportional to
$\rhoDM \sigmaN$, rather than $\rhoDM$ or $\sigmaN$ independently,
where $\rhoDM$ is the local density of dark matter (assumed here to be
dominated by relic neutralinos) and $\sigmaN$ is any of the four
$\chi$-nucleon cross sections.  Since the local dark matter density
has not been directly measured, it is typically inferred from galactic
dynamics and $N$-body simulations.  By convention, experimental results
are often presented for a neutralino density of 0.3~GeV/cm$^3$, with
the implicit understanding that the following significant uncertainties
exist in this value.  In the case of a smooth distribution of galactic
dark matter, $\rhoDM$ is estimated to be
0.2-0.4~GeV/cm$^3$~\cite{BinneyTremaine,Jungman:1995df} for a
spherical halo, but it may be somewhat higher,
up to 0.7~GeV/cm$^3$~\cite{Gates:1995dw,Kamionkowski:1997xg}, for an
elliptical halo; see Ref.~\cite{Jungman:1995df} for a discussion of
the difficulties in determining this value.
Models of the galaxy based
upon hierarchical formation~\cite{Kamionkowski:2008vw}, which do not
assume a strictly smooth distribution of dark matter as do the above
estimates, suggest that the local density may be as low as
0.04~GeV/cm$^3$, but that 0.2~GeV/cm$^3$ is a more reasonable lower
limit.  With this hierachical formation, the halo may contain
substructure such as clumps~\cite{Klypin:1999uc,Moore:1999nt} or tidal
streams from galaxy mergers~\cite{SgrGalaxy} that can increase the
dark matter density.
An analysis of the Sagittarius dwarf galaxy, whose leading tidal tail
passes through the galactic disk somewhere near our location, estimates
it may contribute an additional 0.001-0.07~GeV/cm$^3$ to the local dark
matter density~\cite{SgrStream}.
Without detailed knowledge of the size and prevalence of all such
substructure (particularly, the \textit{locally} present
substructure), it is difficult to place upper limits on the local
dark matter density; however, an increase by greater than a factor of a
few over the above estimates are unlikely~\cite{Kamionkowski:2008vw}.
On the other hand, one should in principle allow for the possibility
that there may be some additional source of cold dark matter, such as
axions, in which case the neutralino density would be less than
$\rhoDM$, though we do not consider this possibility here.
Without a precise determination of $\rhoDM$, it is difficult to place
precise limits on the $\chi$-nucleon cross sections from direct
detection experiments. However, the \textit{ratios} of cross-sections
may be unambiguously determined, so we focus on these ratios in the
following.

\begin{figure*}
  \includegraphics[width=1.00\textwidth]{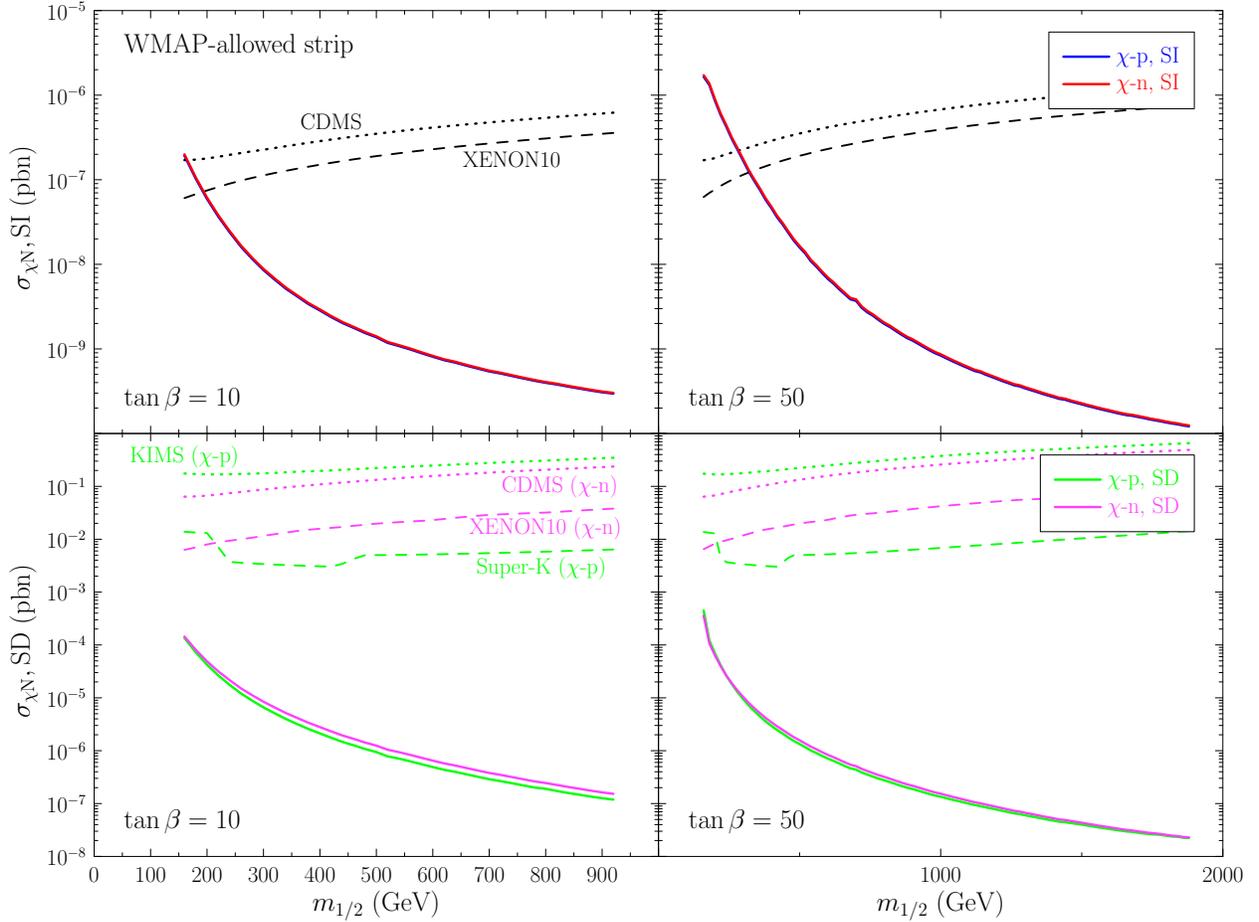}
  \caption[Cross-section (WMAP coannihilation strip)]{
    \coloronlinestatement{}
    The neutralino-nucleon scattering cross sections along the
    WMAP-allowed coannihilation strip for $\tanb = 10$ (left panels)
    and coannihilation/funnel strip for $\tanb = 50$ (right panels)
    using the parameters in \reftab{params}.
    CMSSM predictions for the spin-independent (SI) cross sections
    are shown in the upper panels ($\sigmapSI$ in solid/blue,
    $\sigmanSI$ in solid/red), along with experimental contraints
    from CDMS (dotted/black) and XENON10 (dashed/black),
    where the constraints apply to either SI cross section under the
    reasonable MSSM assumption $\sigmapSI \approx \sigmanSI$.
    Note that $\sigmapSI$ and $\sigmanSI$ are nearly indistinguishable
    at the scale used for these plots.
    The lower panels show CMSSM predictions for the spin-dependent (SD)
    cross-sections ($\sigmapSD$ in solid/green, $\sigmanSD$ in
    solid/magenta).  Experimental constraints for $\sigmapSD$ are
    shown for KIMS (dotted/green) and Super-Kamiokande (dashed/green),
    while constraints for $\sigmanSD$ are given by CDMS
    (dotted/magenta) and XENON10 (dashed/magenta).
    Limits are based upon a local neutralino density of 0.3~GeV/cm$^3$.
    }
  \label{fig:CS}
\end{figure*}

We take the fiducial values of the experimental quantities to be those
obtained using the central values of the parameters in \reftab{params}.
For illustration, cross-sections and their ratios are given in
\reftab{models} for a few benchmark models which are essentially the
well-studied benchmark models C, L, and M~\cite{bench}.  All of the
benchmark models lead to relic densities within the WMAP preferred
range of $\Omega h^2 = 0.088 - 0.120$~\cite{Spergel:2006hy}, and
satisfy most phenomenological constraints~\footnote{The exception being
        a possible failure to account for the discrepancy between
        the theoretical and experimental results for anomalous magnetic 
        moment of the muon~\cite{newBNL,gminus2}.  Points C and L are
        consistent with $(g-2)_\mu$, whilst the contribution from point
        M is too small.}.
Points C and L are points along the coannihilation strip (at low
$\tanb = 10$ and high $\tanb = 50$, respectively) where the masses of
the neutralino and stau are nearly degenerate.  Point M is in the
funnel region where the mass of the neutralino is roughly half that of
the Higgs pseudoscalar.  \Reffig{CS} shows the cross sections along
the WMAP-allowed strips of the $(m_{1/2}, m_0)$ planes for $\tanb$ = 10
and 50~\cite{cmssmwmap,osusy07,eosk2}. For $\tanb = 10$, this is the
coannihilation strip and includes point C, whereas for $\tanb = 50$,
the strip is formed by coannihilations at low $m_{1/2}$, turning
into the rapid annihilation funnel at larger $m_{1/2}$.  This strip
includes points L and M.

Current experimental upper limits on cross sections are also given in
\reffig{CS}.  All these limits assume a local neutralino density of
0.3~GeV/cm$^3$, and should be rescaled by (0.3~GeV/cm$^3$)/$\rhoDM$
for other values of $\rhoDM$.  We display the SI limits
given by CDMS~\cite{Akerib:2005kh} and XENON10~\cite{xenon} under the
reasonable CMSSM assumption that $\sigmapSI \approx \sigmanSI$;
relaxing that assumption would weaken the given limits by at most
a factor of $\sim$5.  The same assumption cannot be made for the
SD cross-sections, and experiments typically constrain only one of
$\sigmapSD$ or $\sigmanSD$, not both.  The direct detection
experiment KIMS~\cite{Lee:2007qn} provides a limit for $\sigmapSD$,
while CDMS~\cite{Akerib:2005za} and XENON10~\cite{xenonSD} (taken
from \cite{DMLPG}) constrain $\sigmanSD$.  Since the Sun is primarily
composed of hydrogen, the capture rate of neutralinos is particularly
sensitive to $\sigmapSD$; thus, indirect detection experiments are
able to constrain this cross-section.  The large mass of the Sun
and the large size of the detector allows
Super-Kamiokande~\cite{Desai:2004pq} to provide a significantly
better limit on $\sigmapSD$ than the direct detection experiments.
It is apparent from \reffig{CS} that experiments are beginning
to probe the CMSSM parameter space through SI scattering, 
particularly at large $\tan \beta$, but
are not yet sensitive enough to detect SD scattering.

\begin{table*}
  \begin{ruledtabular}
  \begin{tabular}{lccc}
    Model            & C    & L    & M    \\
    \hline
    $m_{1/2}$ (GeV)  & 400  & 460  & 1840 \\
    $m_{0}$ (GeV)    & 90   & 310  & 1400 \\
    $\tanb$          & 10   & 50   & 50   \\
    $A_0$            & 0    & 0    & 0    \\
    $\signmu$        & +    & +    & +    \\
    \hline
    $m_{\chi}$ (GeV) & 165  & 193  & 830  \\
    \hline
    $\sigmapSI$ (pb) & $2.85 \times 10^{-9}$
                     & $2.36 \times 10^{-8}$
                     & $1.28 \times 10^{-10}$ \\
    $\quad$68.3\% C.L. & $(1.65 - 4.47) \times 10^{-9}$
                       & $(1.23 - 3.95) \times 10^{-8}$
                       & $(0.76 - 1.98) \times 10^{-10}$ \\
    $\quad$95.4\% C.L. & $(0.81 - 6.46) \times 10^{-9}$
                       & $(0.49 - 5.98) \times 10^{-8}$
                       & $(0.39 - 2.83) \times 10^{-10}$ \\
    $\sigmanSI$ (pb) & $2.93 \times 10^{-9}$
                     & $2.46 \times 10^{-8}$
                     & $1.32 \times 10^{-10}$ \\
    $\quad$68.3\% C.L. & $(1.72 - 4.56) \times 10^{-9}$
                       & $(1.31 - 4.07) \times 10^{-8}$
                       & $(0.79 - 2.02) \times 10^{-10}$ \\
    $\quad$95.4\% C.L. & $(0.86 - 6.58) \times 10^{-9}$
                       & $(0.55 - 6.10) \times 10^{-8}$
                       & $(0.41 - 2.87) \times 10^{-10}$ \\
    $\sigmapSD$ (pb) & $2.19 \times 10^{-6}$
                     & $1.82 \times 10^{-6}$
                     & $2.40 \times 10^{-8}$ \\
    $\quad$68.3\% C.L. & $(1.91 - 2.49) \times 10^{-6}$
                       & $(1.62 - 2.04) \times 10^{-6}$
                       & $(2.19 - 2.63) \times 10^{-8}$ \\
    $\quad$95.4\% C.L. & $(1.64 - 2.81) \times 10^{-6}$
                       & $(1.43 - 2.26) \times 10^{-6}$
                       & $(1.98 - 2.86) \times 10^{-8}$ \\
    $\sigmanSD$ (pb) & $2.81 \times 10^{-6}$
                     & $2.10 \times 10^{-6}$
                     & $2.45 \times 10^{-8}$ \\
    $\quad$68.3\% C.L. & $(2.49 - 3.14) \times 10^{-6}$
                       & $(1.89 - 2.33) \times 10^{-6}$
                       & $(2.23 - 2.67) \times 10^{-8}$ \\
    $\quad$95.4\% C.L. & $(2.18 - 3.51) \times 10^{-6}$
                       & $(1.68 - 2.57) \times 10^{-6}$
                       & $(2.03 - 2.91) \times 10^{-8}$ \\
    \hline
    $\sigmanSI / \sigmapSI$ & $1.029$ & $1.042$ & $1.026$ \\
    $\quad$68.3\% C.L.  & $1.020 - 1.042$
                        & $1.028 - 1.065$
                        & $1.018 - 1.037$ \\
    $\quad$95.4\% C.L.  & $1.015 - 1.066$
                        & $1.020 - 1.114$
                        & $1.013 - 1.056$ \\
    $\sigmanSD / \sigmapSD$ & $1.28$  & $1.15$  & $1.02$  \\
    $\quad$68.3\% C.L.  & $1.00 - 1.65$
                        & $0.93 - 1.44$
                        & $0.85 - 1.22$ \\
    $\quad$95.4\% C.L.  & $0.78 - 2.14$
                        & $0.75 - 1.80$
                        & $0.71 - 1.46$ \\
    $\sigmapSD / \sigmapSI$ & $770$   & $77$  & $187$   \\
    $\quad$68.3\% C.L.  & $480 - 1350$
                        & $46 - 151$
                        & $121 - 319$ \\
    $\quad$95.4\% C.L.  & $320 - 2730$
                        & $30 - 373$
                        & $83 - 616$ \\
    $\sigmanSD / \sigmanSI$ & $960$   & $86$  & $186$   \\
    $\quad$68.3\% C.L.  & $610 - 1660$
                        & $51 - 163$
                        & $120 - 313$ \\
    $\quad$95.4\% C.L.  & $410 - 3370$
                        & $33 - 392$
                        & $83 - 601$ \\
  \end{tabular}
  \end{ruledtabular}
  \caption[Parameters]{
    Neutralino-nucleon scattering cross sections and ratios for
    benchmark models C, L, and M.  Confidence intervals are given
    for confidence levels (C.L.) of 68.3\% and 95.4\%, using the
    hadronic parameter uncertainties in \reftab{params}.
    }
  \label{tab:models}
\end{table*}

From our analysis of the benchmark models and along the WMAP
strip, several general features of the $\chi$-nucleon cross sections
are apparent for any given CMSSM model.  The SI cross sections for
$\chi$-proton ($\sigmapSI$) and $\chi$-neutron ($\sigmanSI$) scattering
are typically very close, within $\sim$5\% of each other; close enough
that they are virtually indistinguishable in \reffig{CS}.
Likewise, the spin-dependent $\chi$-proton ($\sigmapSD$) and
$\chi$-neutron ($\sigmanSD$) scattering cross sections are similar,
differing by at most a factor of 2 or 3.  In general, the SD
$\chi$-nucleon cross section is much larger than the SI one, by
$\orderof{10^2-10^3}$ or more.  However, we recall that the SI
cross section for scattering off a nucleus, \refeqn{sigmaSI}, contains
a factor of the number of nucleons squared, whereas the SD
cross section, \refeqn{sigmaSD}, is proportional to the square of the
spin, which does not grow with increasing nuclear mass. Consequently,
heavy elements such as Ge and Xe are actually more sensitive to SI
couplings than to SD couplings.

\subsection{\label{sec:CI} Confidence Intervals}

To determine the level of uncertainty in the cross sections and ratios
obtained for a given model, we generate confidence intervals about the
central values of the resulting cross section and ratio distributions
by Monte Carlo sampling over the parameters in \reftab{params},
assuming independent Gaussian errors~\footnote{The distributions for
        the cross sections generated by the Monte Carlo are not even
        approximately Gaussian or symmetric as is evidenced by the
        curvature seen in \Reffig{BenchmarkCS}.  Hence, confidence
        intervals are used instead of the averages and standard
        deviations of the generated distributions.}.
Results are given in \reftab{models} for benchmark models C, L, and M
at confidence levels (C.L.) of 68.3\% and 95.4\%.
These confidence intervals take into account only hadronic
uncertainties, not any uncertainties arising within the supersymmetric
models themselves.  We recall, in particular, that the estimates for
the $\alpha_{2q}$ and $\alpha_{3q}$ coefficients of the effective
Lagrangian in \refeqn{lagrangian} contain uncertainties that may only
be reduced by calculating radiative corrections to the effective
scattering operators, work that is beyond the scope of this paper.
These confidence intervals take into account only the explicit
dependence of $\mt$ in Eqns.~(\ref{eqn:alpha2}), (\ref{eqn:alpha3}),
\&~(\ref{eqn:fN}), but not in MSSM terms such as $\theta_{\rm t}$,
$m_A$, and $|\mu|$, which are calculated using the central value of
171.4~GeV.  The confidence intervals are not significantly affected by
this incomplete treatment of the top quark mass uncertainty, apart from
a slight underestimate of the $\sigmaNSD$ interval widths; we shall
address below the full MSSM dependence on the top quark mass.

As indicated by the benchmark models, the hadronic uncertainties
induce significant uncertainties in the cross sections.  At the 68.3\%
C.L., the SI cross sections vary by $\sim$3; at the 95.4\% C.L., they
vary by an order of magnitude.  Such a variation is larger than
might na\"\i{}vely be expected given the much smaller relative
uncertainties of the parameters listed in \reftab{params}, the
largest of which is $\md$ at 5$\pm$2~MeV~\footnote{We show below that,
        whilst having the largest relative uncertainty, $\md$ makes a
        negligible contribution to the uncertainties in the cross
        sections.}.
Clearly, such a large uncertainty in the cross sections would
make it difficult to narrow down CMSSM parameters on the basis of any
single experimental signal.

Whilst the two SI cross sections can vary greatly, their ratio does not:
in our Monte Carlo analysis, $\sigmanSI / \sigmapSI$ varies by only a
few percent.  It is worth noting, however, that in some models
(\eg\ model L), $\sigmanSI$ may exceed $\sigmapSI$ by more than 10\%
at the 95.4\% C.L., a difference that may be of experimental interest.
In addition, even at the 95.4\% C.L., $\sigmanSI$ is always larger than
$\sigmapSI$.

The variations in the two SD cross sections are not so large:
$\sim$20\% and $\sim$40\% at the 68.3\% and 95.4\% C.L.'s, respectively.
The ratio of the two cross sections, however, has a larger variation,
typically by a factor of 2 or 3, due to an anti-correlation in the
variations of the two cross sections that we discuss below.

\begin{table}
  \begin{ruledtabular}
  \begin{tabular}{lcccc}
    & $\sigmapSI$ (pb) & $\sigmapSD$ (pb)
      & \ifmulticol{$\displaystyle\frac{\sigmanSI}{\sigmapSI}$}%
                   {$\sigmanSI / \sigmapSI$}
      & \ifmulticol{$\displaystyle\frac{\sigmanSD}{\sigmapSD}$}%
                   {$\sigmanSD / \sigmapSD$}
      \\
    \hline
    fiducial value & $2.85 \times 10^{-9}$ & $2.19 \times 10^{-6}$
                   & $1.029$               & $1.28$ \\
    \hline
    \mumd & $\pm 3.5\%$               & $\sim 0.$%
            \makebox[0pt][l]{\footnotemark[1]}
          & $\pm 0.08\%$              & $\sim 0.$ \\
    \md   & $\sim 0.$                 & $\sim 0.$
          & $\sim 0.$                 & $\sim 0.$ \\
    \msmd & $\pm 5.2\%$               & $\sim 0.$
          & $\pm 0.07\%$              & $\sim 0.$ \\
    \mc   & $\sim 0.$                 & $\sim 0.$
          & $\sim 0.$                 & $\sim 0.$ \\
    \mb   & $\pm 0.1\%$               & $\pm 0.04\%$
          & $\sim 0.$                 & $\pm 0.01\%$ \\
    \mt   & $\pm 9\%$                 & $\pm 9\%$
          & $\pm 0.005\%$             & $\pm 2\%$ \\
    \hline
    $\sigma_0$ & $_{-27\%}^{+34\%}$              & --
               & $_{-0.7\%}^{+1.2\%}$            & -- \\[1.0ex]
    \SigmapiN  & $_{-32\%}^{+45\%}$              & --
               & $_{-0.4\%}^{+0.7\%}$            & -- \\
    \hline
    \athree  & -- & $\pm 0.5\%$
             & -- & $\pm 0.06\%$ \\
    \aeight  & -- & $\pm 2.2\%$
             & -- & $\pm 4.2\%$ \\
    \Deltaps & -- & $_{-12\%}^{+14\%}$
             & -- & $_{-21\%}^{+30\%}$ \\
  \end{tabular}
  \end{ruledtabular}
  \footnotetext[1]{Entries listed as $\sim 0.$ have a relative
                   variation less than $10^{-5}$ 
                   that of the fiducial value.}
  \caption[Uncertainties]{
    Relative uncertainties in $\sigmapSI$, $\sigmapSD$,
    $\sigmanSI / \sigmapSI$, and $\sigmanSD / \sigmapSD$ for Model C
    due to each of the parameters in \reftab{params}.
    The quoted uncertainties correspond to 68.3\% C.L.\ confidence
    intervals relative to the fiducial values.
    Uncertainties in $\sigmanSI$ and $\sigmanSD$ (not shown) are
    comparable to those in $\sigmapSI$ and $\sigmapSD$, respectively.
    Due to the non-linear dependence of the cross sections on some of
    the parameters over the ranges of interest,
    some confidence intervals are not symmetric about the fiducial
    values.
    }
  \label{tab:errors}
\end{table}

To demonstrate the contributions of individual parameters to the
cross-section uncertainties, we have determined 68.3\% C.L.\ confidence
intervals for variations in the parameters treated singly, with the
remaining parameters fixed at their central values, as presented in
\reftab{errors} for benchmark model C.  The intervals are given as
variations relative to the fiducial values.  For several parameters,
notably $\sigma_0$, $\SigmapiN$, and $\Deltaps$, the variations in
the cross sections are non-linear with respect to the parameters over
the values of interest, resulting in confidence intervals that are
not symmetric about the fiducial values.  The variations due to the
top mass here include the dependence of MSSM parameters such as
$\theta_{\rm t}$, $A_0$, and $|\mu|$ on $\mt$, which was neglected
in the full confidence intervals of \reftab{models}.

It is clear that from \reftab{errors} that the uncertainties in the
quark masses and their ratios, apart from the top mass, make almost
negligible contributions to the cross-section uncertainties.
The SI cross-section uncertainties are dominated by contributions
from $\sigma_0$ and $\SigmapiN$, and the SD cross-section
uncertainties mainly arise from uncertainties in $\Deltaps$.  We
examine these in the following sections.

\subsection{\label{sec:mq} Quark Masses and Mass Ratios}

The most significant quark mass dependence of the $\chi$-nucleon
scattering cross sections is that on the top-quark mass, which induces
uncertainties of $\sim$10\% in all four $\sigmaN$.  These
uncertainties arise not from the appearance of $\mt$ in 
Eqns.~(\ref{eqn:alpha2}), (\ref{eqn:alpha3}), \&~(\ref{eqn:fN})
(hereafter referred to as the explicit $\mt$ dependence), but in the
calculations of CMSSM parameters, \eg, $m_A$ and $|\mu|$.  However,
variations in $\mt$ rescale all four cross-sections in the same manner, 
so that ratios of the cross-sections are essentially independent of
$\mt$.  The precision of the top mass measurement will continue to
improve once the LHC begins taking data, with a 1~GeV uncertainty a
possibility in as little as one year of low-luminosity 
running~\cite{Borjanovic:2004ce}.  Thus, we can expect the $\mt$-induced
uncertainties in $\sigmaN$ to fall from $\sim$10\% to $\sim$5\% or
lower within a few years, well below uncertainties induced by other
parameters.

The SI cross sections are sensitive to the ratios of the light quark
masses $\mup/\md$ and $\ms/\md$, more than to the overall scale of the
light quark masses, which is fixed by the relatively poorly constrained
value of $\md$.
In the CMSSM, $\theta_f \approx 0$ for the light quarks ($u,d,s$) and
the charm quark ($c$), so that
$\eta$ in \refeqn{etamatrix} is nearly diagonal.  For diagonal $\eta$,
$\alpha_{3q}$ is proportional to $m_{q}$, and hence
$\alpha_{3q}/m_{q}$ in \refeqn{fN} is independent of the quark mass.
Any dependence on the light quark masses must come through
the $f_{Tq}^{(\rmpp,\rmnn)}$ terms.  However, by
Eqns.~(\ref{eqn:fNTu})-(\ref{eqn:fNTs}), those terms depend only on
mass ratios.  Since there is no dependence on $\md$ (for fixed mass
ratios), its relatively large uncertainty does not
translate into any significant uncertainties in the SI cross sections.

Whilst the mass ratios induce uncertainties of a few \% in the
cross-sections, the induced uncertainties in the ratio
$\sigmanSI/\sigmapSI$ are only $\sim$0.1\%.  This is a consequence of
the fact that the predominant contribution to \refeqn{fN} comes from
the strange-quark term $f_{T\rms}^{(\rmpp,\rmnn)} \alpha_{3\rms}
/ m_{\rms}$, which is common to both $f_{\rmpp}$ and $f_{\rmnn}$.
Thus, variations in the the quark mass ratios induce, via \refeqn{fNTs},
comparable (and correlated) variations in both $\sigmapSI$ and
$\sigmanSI$, resulting in only small variations in the ratio of those
two cross sections.

The $\eta$ matrices for the bottom and top quarks, unlike the lighter
quarks, are not approximately diagonal, so that $\alpha_{3q}/m_{q}$
does have a quark mass dependence. However, the precisions to which
these quark masses are known lead to only small uncertainties
($<$0.1\%) in the SI cross-sections (here, we refer only to the
explicit $\mt$ dependence in these terms, not to the MSSM parameter
dependence discussed previously).

The only quark mass dependence of the SD cross sections in
\refeqn{aN} comes from the light quark masses in $\alpha_{2q}$
($q = u,d,s$).  However, mass-dependent terms are suppressed
by a factor of $\orderof{m_q/m_W}$ ($\orderof{m_q^2/m_W^2}$ for
diagonal $\eta$ matrices) relative to mass-independent terms
arising from the $X_i$, $Y_i$, $W_i$, and $V_i$ factors given by
\refeqn{XYWV}. Thus, the SD cross sections are nearly independent of
the quark masses.

\subsection{\label{sec:SI} Spin-Independent Parameters and
            Cross Sections}

\begin{table}
  \begin{ruledtabular}
  \begin{tabular}{lccc}
    Model & C & L & M    \\
    \hline
    $\SigmapiN$ = 36 MeV: & & & \\
    \hspace{2em}$\sigmapSI$ (pb) & 3.40 $\times 10^{-10}$
                                 & 1.38 $\times 10^{-9}$
                                 & 1.78 $\times 10^{-11}$ \\
    \hspace{2em}$\sigmanSI$ (pb) & 3.67 $\times 10^{-10}$
                                 & 1.61 $\times 10^{-9}$
                                 & 1.89 $\times 10^{-11}$ \\
    \hspace{2em}$\sigmanSI / \sigmapSI$ & 1.080
                                        & 1.170
                                        & 1.065 \\
    $\SigmapiN$ = 45 MeV: & & & \\
    \hspace{2em}$\sigmapSI$ (pb) & 8.80 $\times 10^{-10}$
                                 & 5.55 $\times 10^{-9}$
                                 & 4.23 $\times 10^{-11}$ \\
    \hspace{2em}$\sigmanSI$ (pb) & 9.24 $\times 10^{-10}$
                                 & 6.02 $\times 10^{-9}$
                                 & 4.41 $\times 10^{-11}$ \\
    \hspace{2em}$\sigmanSI / \sigmapSI$ & 1.050
                                        & 1.085
                                        & 1.043 \\
    $\SigmapiN$ = 56 MeV: & & & \\
    \hspace{2em}$\sigmapSI$ (pb) & 1.88 $\times 10^{-9}$
                                 & 1.45 $\times 10^{-8}$
                                 & 8.64 $\times 10^{-11}$ \\
    \hspace{2em}$\sigmanSI$ (pb) & 1.95 $\times 10^{-9}$
                                 & 1.52 $\times 10^{-8}$
                                 & 8.91 $\times 10^{-11}$ \\
    \hspace{2em}$\sigmanSI / \sigmapSI$ & 1.035
                                        & 1.053
                                        & 1.031 \\
    $\SigmapiN$ = 64 MeV: & & & \\
    \hspace{2em}$\sigmapSI$ (pb) & 2.85 $\times 10^{-9}$
                                 & 2.36 $\times 10^{-8}$
                                 & 1.28 $\times 10^{-10}$ \\
    \hspace{2em}$\sigmanSI$ (pb) & 2.93 $\times 10^{-9}$
                                 & 2.46 $\times 10^{-8}$
                                 & 1.32 $\times 10^{-10}$ \\
    \hspace{2em}$\sigmanSI / \sigmapSI$ & 1.029
                                        & 1.042
                                        & 1.026 \\
    $\SigmapiN$ = 72 MeV: & & & \\
    \hspace{2em}$\sigmapSI$ (pb) & 4.01 $\times 10^{-9}$
                                 & 3.49 $\times 10^{-8}$
                                 & 1.78 $\times 10^{-10}$ \\
    \hspace{2em}$\sigmanSI$ (pb) & 4.11 $\times 10^{-9}$
                                 & 3.61 $\times 10^{-8}$
                                 & 1.82 $\times 10^{-10}$ \\
    \hspace{2em}$\sigmanSI / \sigmapSI$ & 1.025
                                        & 1.035
                                        & 1.022 \\
    $\SigmapiN$ = 84 MeV: & & & \\
    \hspace{2em}$\sigmapSI$ (pb) & 6.13 $\times 10^{-9}$
                                 & 5.61 $\times 10^{-8}$
                                 & 2.69 $\times 10^{-10}$ \\
    \hspace{2em}$\sigmanSI$ (pb) & 6.26 $\times 10^{-9}$
                                 & 5.76 $\times 10^{-8}$
                                 & 2.74 $\times 10^{-10}$ \\
    \hspace{2em}$\sigmanSI / \sigmapSI$ & 1.021
                                        & 1.028
                                        & 1.019 \\
  \end{tabular}
  \end{ruledtabular}
  \caption[$\SigmapiN$]{
    Spin-independent neutralino-nucleon scattering
    cross sections in the benchmark models for several values of
    $\SigmapiN$.
    }
  \label{tab:Sigma}
\end{table}

The greatest impediment to an accurate determination of the SI
cross sections for any given MSSM model comes from the $\SigmapiN$
and $\sigma_0$ parameters.  As shown for benchmark model C in
\reftab{errors}, each of these two parameters induces uncertainties
of $\sim$30\% or more in $\sigmapSI$ at the 68.3\% C.L.; $\sigmanSI$
(not shown) has similar induced uncertainties.
The large confidence intervals for the SI cross sections in
\reftab{models} are almost entirely due to the uncertainties in these
two parameters.

We focus the discussion here mainly on $\SigmapiN$ rather than
$\sigma_0$ as the $\SigmapiN = 64 \pm 8$ MeV result is significantly
larger than previous estimates for that parameter. Indeed the range of 
estimates of the central value for $\SigmapiN$ is far greater than the
typically quoted uncertainty.  In view of this, we also
include below some results for lower values of $\SigmapiN$.

\begin{figure}
  \includegraphics[width=\scfigwidth]{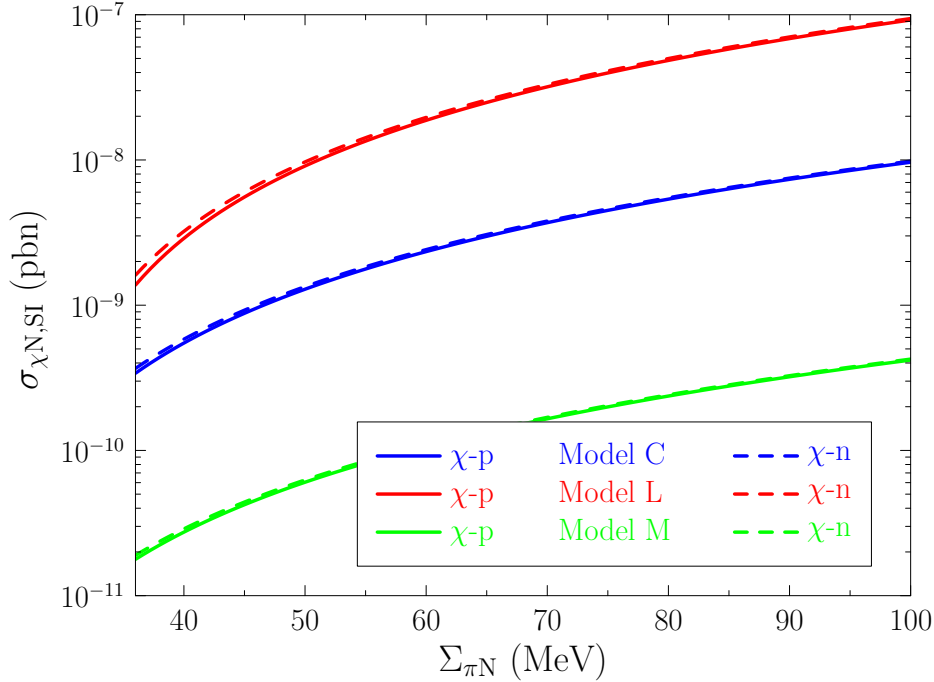}
  \caption[Cross-section vs.\ $\SigmapiN$]{
    \coloronlinestatement{}
    The spin-independent neutralino-nucleon scattering cross section
    as a function of $\SigmapiN$ for benchmark models C, L, and M.
    Note that $\sigmapSI$ and $\sigmanSI$ are nearly indistinguishable
    at the scale used in this plot.
    }
  \label{fig:BenchmarkCS}
\end{figure}

\begin{figure*}
  \includegraphics[width=1.00\textwidth]{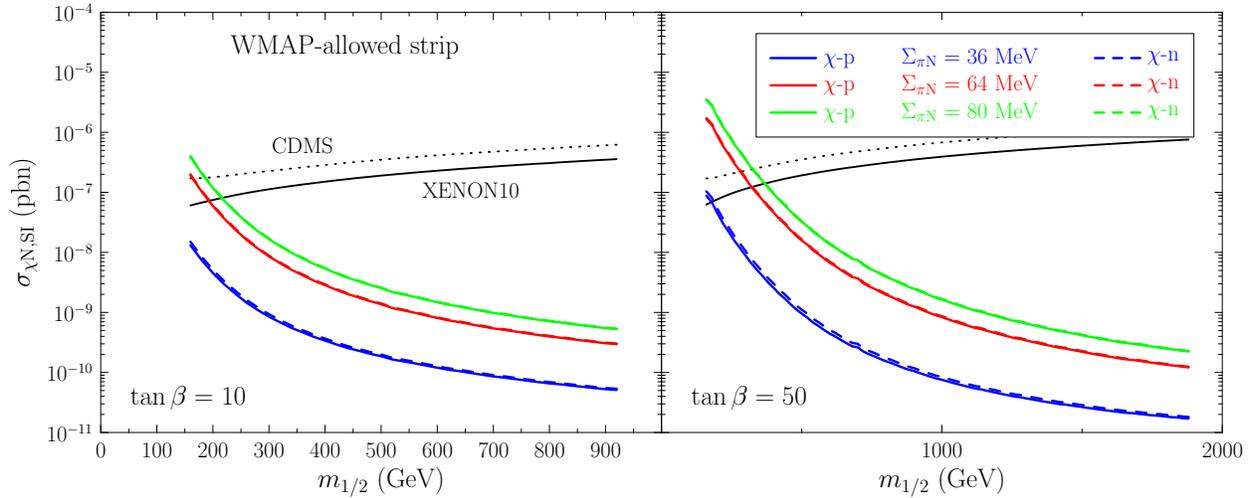}
  \caption[The SI cross section for several values of $\SigmapiN$
           (WMAP strip)]{
    \coloronlinestatement{}
    The spin-independent neutralino-nucleon scattering cross section
    ratio along the WMAP-allowed strips for $\tanb = 10$
    (left panel) and $\tanb = 50$ (right panel) for several values of
    $\SigmapiN$.
    Experimental constraints from CDMS and XENON10 are also shown.
    }
  \label{fig:CSSigma}
\end{figure*}

In \reffig{BenchmarkCS}, we show the $\SigmapiN$ dependence of
$\sigmaNSI$ for the benchmark models, and \reftab{Sigma} gives
the $\sigmaNSI$ values for those models for selected values of
$\SigmapiN$.  All the other parameters are set at their fiducial values
(\reftab{params}).
From the minimal value for $\SigmapiN$ ($\sigma_0$ = 36~MeV) to the
2-$\sigma$ upper bound (80~MeV), $\sigmaNSI$ varies by more than a
factor of 10 (as much as a factor of 35 for model L).
At these benchmarks and in other models of interest, for larger values
of $\SigmapiN$ ($\SigmapiN \not\sim \sigma_0$), the majority of the
contribution to $f_{\rmpp}$ in \refeqn{fN} comes from the strange
quark term, with $\fpTq{s} \propto y \propto \SigmapiN - \sigma_0$,
so that $\sigmapSI \sim (\SigmapiN - \sigma_0)^2$.
Thus, the SI cross sections are particularly sensitive not just to
$\SigmapiN$ and $\sigma_0$, but to their difference.
For smaller values of $\SigmapiN$ ($\SigmapiN \sim \sigma_0$), the
strange contribution no longer dominates, but a strong dependence of
$\sigmaNSI$ on $\SigmapiN$ does remain.

In \reffig{CSSigma}, the SI cross sections are shown along the WMAP
allowed coannihilation strip for $\tanb$ = 10 and
coannihilation/funnel strip for $\tanb = 50$ for the $\SigmapiN$
reference values of 36~MeV (no strange scalar contribution),
64~MeV (central value), and 80~MeV (2-$\sigma$ upper bound);
CDMS and XENON10 limits are also given.  As with the benchmark models,
a factor of $\sim$10 variation occurs in $\sigmaNSI$ over these
$\SigmapiN$ reference values for any given model along the WMAP strip.

Such large variations present difficulties in using any upper limit
or possible future precision measurement of $\sigmaNSI$ from a direct
detection signal to constrain the CMSSM parameters.  The present
CDMS and XENON10 upper limits have (almost) no impact on the WMAP
strip for $\tanb = 10 (50)$, if one makes the very conservative
assumption that $\sigma_0$ = 36~MeV ($y$ = 0). On the other hand,
$m_{1/2} \sim 200$~GeV would be excluded for $\tanb = 10$ if
$\SigmapiN = 64$ or 80~MeV. This excluded region would extend to
$m_{1/2} \sim 300$~GeV for $\tanb = 50$ if $\SigmapiN = 64$ or 80~MeV.
Thus, {\it the experimental uncertainty in $\SigmapiN$ is already
impinging on the ability of the present CDMS and XENON10 results to
constrain the CMSSM parameter space.}

Looking to the future, a conjectural future measurement of
$\sigmapSI = 4 \times 10^{-9}$~pb would only constrain $m_{1/2}$ to the
range 600~GeV $< m_{1/2} <$ 925~GeV if $\tanb = 10$ and 1100~GeV
$< m_{1/2} <$ 1400~GeV if $\tanb = 50$, for the 1-$\sigma$ $\SigmapiN$
range of 64 $\pm$~8~MeV.  If smaller values of $\SigmapiN$ are also
considered, down to $\sigma_0$ = 36~MeV ($y$ = 0), these constraints
would weaken to 350~GeV $< m_{1/2} <$ 925~GeV and 550~GeV $< m_{1/2} <$
1400~GeV for $\tanb$ = 10 and 50, respectively, ruling out only the
smallest values of $m_{1/2}$~\footnote{Moreover, as noted previously,
        however, detection signals only measure $\rhoDM \sigmaN$, so
        $\sigmaN$ can only be determined from a signal to the precision
        that the local dark matter density is known.}.

\begin{figure}
  \includegraphics[width=\scfigwidth]{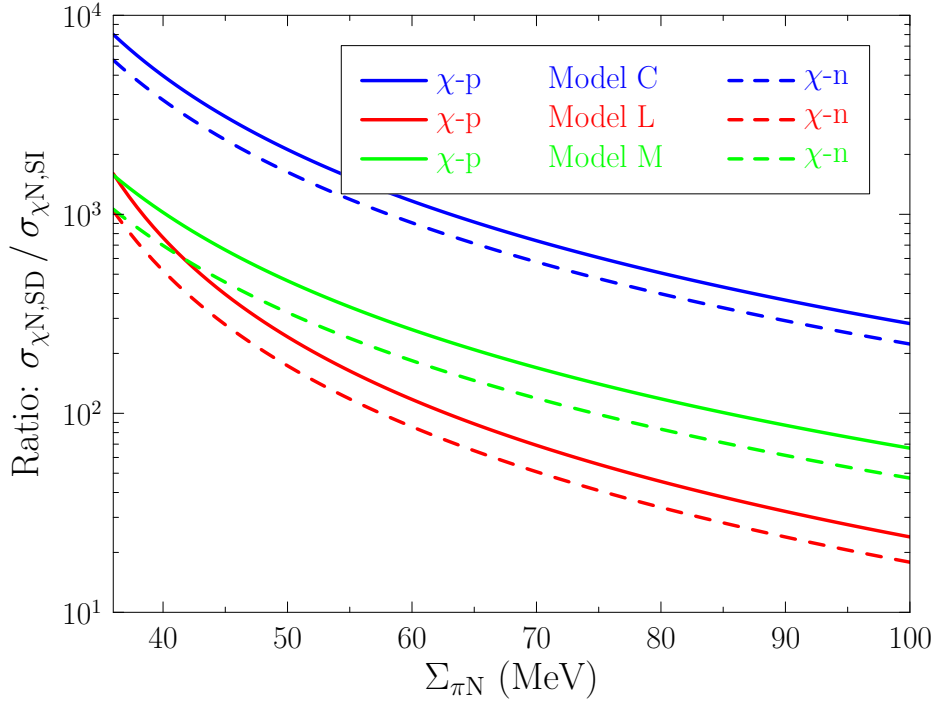}
  \caption[SDSI cross-section ratio vs.\ $\SigmapiN$]{
    \coloronlinestatement{}
    The ratios of the spin-dependent and spin-independent
    neutralino-nucleon scattering cross sections as functions of
    $\SigmapiN$ for the benchmark models C, L, and M.
    }
  \label{fig:BenchmarkSDSIratio}
\end{figure}

\begin{figure*}
  \includegraphics[width=1.00\textwidth]{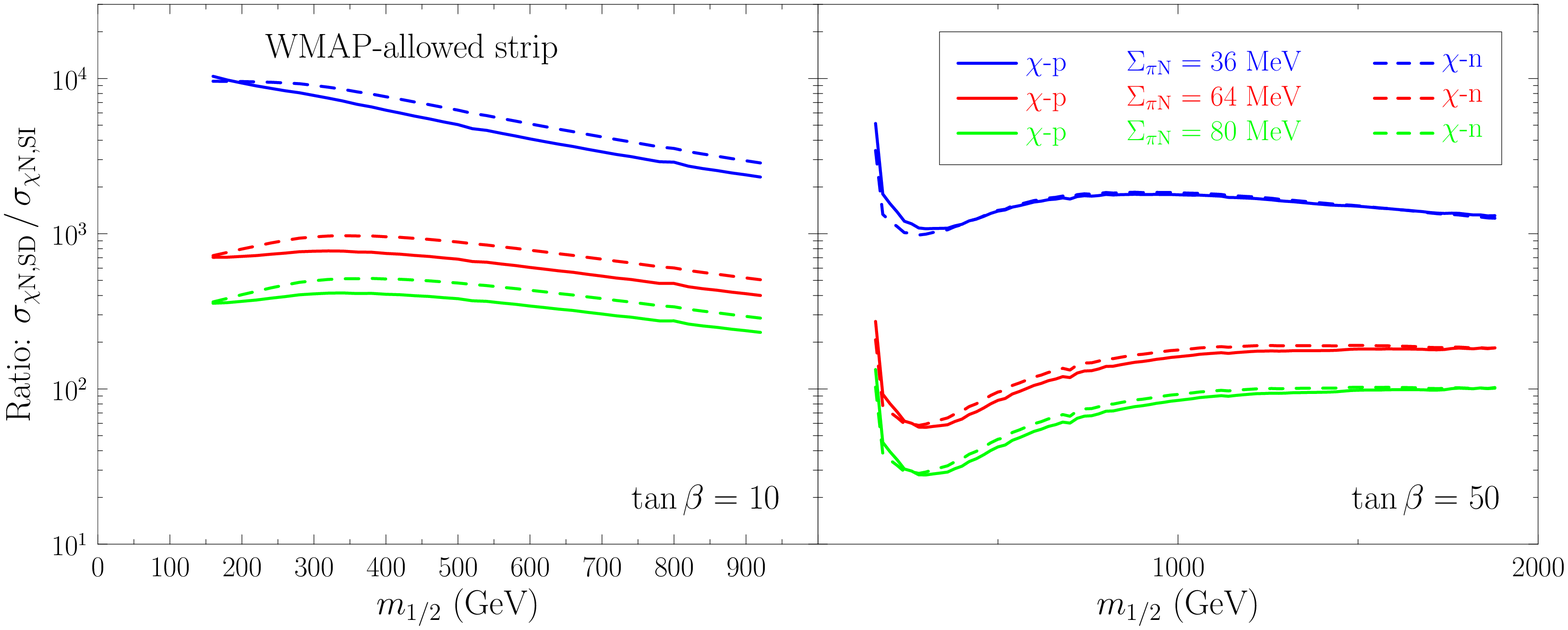}
  \caption[SD/SI cross-section ratio for several values of
           $\SigmapiN$ (WMAP strip)]{
    \coloronlinestatement{}
    The ratios of the spin-dependent and spin-independent
    neutralino-nucleon scattering cross sections along the
    WMAP-allowed strips for $\tanb = 10$ (left panel) and
    $\tanb = 50$ (right panel) for several values of $\SigmapiN$.
    }
  \label{fig:SDSIratioSigma}
\end{figure*}

As the SD cross sections are independent of $\SigmapiN$, the SD/SI
cross-section ratio exhibits the same behavior with respect to
$\SigmapiN$ as the SI cross-section does (albeit inverted), as shown
in \reffig{BenchmarkSDSIratio}.  The ratio is also shown along the
corresponding WMAP  strips in \reffig{SDSIratioSigma} for several
values of $\SigmapiN$~\footnote{The non-monotonic dependence on
        $m_{1/2}$ seen in the right panel of \reffig{SDSIratioSigma}
        for $\tanb = 50$ is due to an enhancement of the Higgsino
        components, $Z_{\chi_3}$ and $Z_{\chi4}$, at low $m_{1/2}$. As
        a result, the $Z$-exchange contribution to the spin-dependent 
        cross section increases rapidly.}.
For small $\SigmapiN$, the ratio is large,
almost $10^4$ for model C and $2000$ for models L and M at
$\SigmapiN$ = 36~MeV, suggesting that spin-sensitive experiments have
a better chance of detecting neutralinos than spin-insensitive
experiments.  At the 2-$\sigma$ upper limit on $\SigmapiN$, however,
the ratio falls to 500 for model C, 50 for model L, and 120 for
model M.  Recalling that the SI $\chi$-nuclear scattering cross section
scales as the number of nucleons squared, SI scattering may actually
dominate over SD scattering for heavier nuclei in the cases of these
lower SD/SI cross-section ratios.

\begin{figure}
  \includegraphics[width=\scfigwidth]{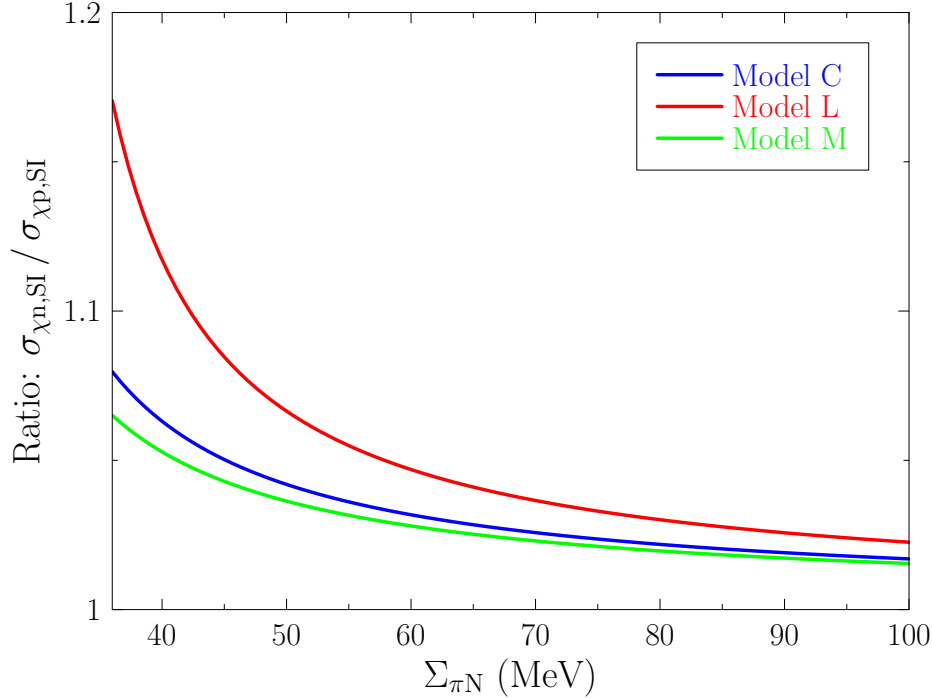}
  \caption[$\chi$-n/$\chi$-p cross-section ratio vs.\ $\SigmapiN$]{
    \coloronlinestatement{}
    The ratios of the spin-independent neutralino-neutron and
    neutralino-proton scattering cross sections as functions of
    $\SigmapiN$ for benchmark models C, L, and M.
    }
  \label{fig:Benchmarknpratio}
\end{figure}

\begin{figure*}
  \includegraphics[width=1.00\textwidth]{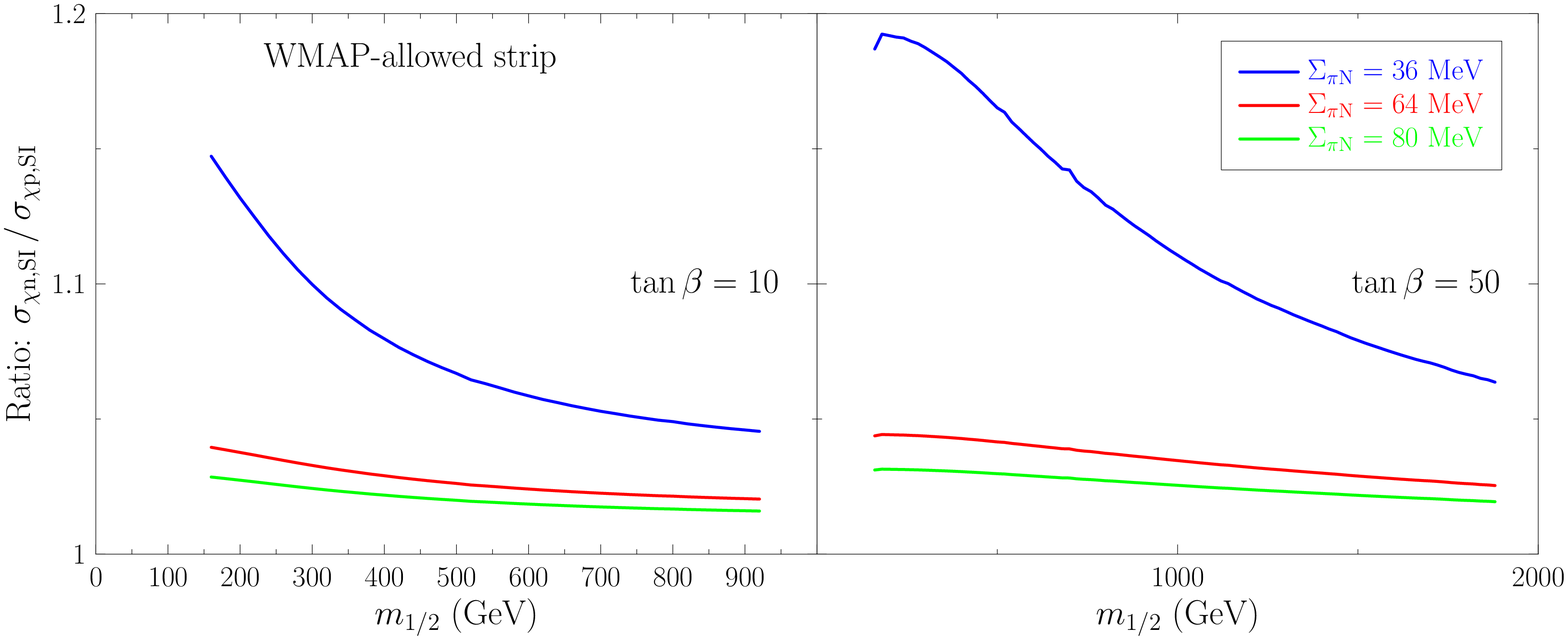}
  \caption[$\chi$-n/$\chi$-p cross-section ratio for several values of
           $\SigmapiN$ (WMAP strip)]{
    \coloronlinestatement{}
    The ratios of the spin-independent neutralino-neutron and
    neutralino-proton scattering cross sections along the
    WMAP-allowed strips for $\tanb = 10$ (left panel) and
    $\tanb = 50$ (right panel) for several values of $\SigmapiN$.
    }
  \label{fig:npratioSigma}
\end{figure*}

The dependence of the SI $\chi$-n/$\chi$-p cross-section ratio on
$\SigmapiN$ is shown in \reffig{Benchmarknpratio} for the benchmark
models, with ratios at selected $\SigmapiN$ given in \reftab{Sigma}.
The $\sigmanSI / \sigmapSI$ ratio is shown along the WMAP
coannihilation strips in \reffig{npratioSigma} for several values
of $\SigmapiN$.
Whilst $\sigmapSI$ and $\sigmanSI$ each depend significantly
on the value of $\SigmapiN$, their ratio has only a mild $\SigmapiN$
dependence.  As noted earlier in this section, the strange-quark
term dominates in \refeqn{fN} for $\SigmapiN \not\sim \sigma_0$ and,
since this term contributes identically to both $f_{\rmpp}$ and
$f_{\rmnn}$ (neglecting nucleon mass differences),
$\sigmanSI / \sigmapSI \approx 1$, independent of $\SigmapiN$.
For $\SigmapiN \sim \sigma_0$, where the strange contribution is no
longer significant, we generally find that
$\fTq{\rmu}, \fTq{\rmd} \ll 1$ so that $f_{TG} \approx 1$ via
\refeqn{fTG} and the right (heavy-quark) summation term in
\refeqn{fN} dominates over the left (light-quark) summation term.
Since the heavy-quark terms are identical in $f_{\rmpp}$ and
$f_{\rmnn}$, we again have $\sigmanSI / \sigmapSI \approx 1$.

The $\fNTq{\rmu}$ and $\fNTq{\rmd}$ terms are small but not entirely
negligible, and the fact that $\fpTq{q} \neq \fnTq{q}$ for $q$=u,d
implies that $f_{\rmpp} \neq f_{\rmnn}$ in general.  The result is a
small shift in the SI $\chi$-n/$\chi$-p cross-section ratio away from
unity.  The shift becomes greater as $\SigmapiN \to \sigma_0$, since
the dominant strange-quark contribution disappears and allows the
proton/neutron asymmetry in the smaller up and down quark terms to
become relevant.  For the benchmark models in \reffig{Benchmarknpratio}
and along the WMAP coannihilation strips in \reffig{npratioSigma},
$\sigmanSI$ is $\sim$5\% larger than $\sigmapSI$ in the 2-$\sigma$
$\SigmapiN$ range $64 \pm 16$~MeV.  For $\SigmapiN \sim \sigma_0$,
however, the difference can become as large as 15 to 20 \%.

It can be shown that $\sigmanSI / \sigmapSI > 1$ for
$|\alpha_{3\rmd}| > |\alpha_{3\rmu}|$.  Since
$\alpha_{3\rmu} \propto 1/B_{\rmu} = 1/\sin{\beta}$ and
$\alpha_{3\rmd} \propto 1/B_{\rmd} = 1/\cos{\beta}$, that
condition is generally satisfied for the $\tanb \gg 1$ models
examined here.  Hence, the neutron always has a slightly larger SI
cross section than the proton in these models.

When including all the uncertainties, as shown by the Monte Carlo
results in \reftab{models}, a small but non-zero asymmetry in the
proton and neutron SI cross-sections is expected to occur at the
$\sim$ few \% level.  Such an asymmetry may be observable, but a
problem arises in the current generation of experiments.  To
extract the relative strength of the $f_{\rmpp}$ and $f_{\rmnn}$
couplings in \refeqn{sigmaSI}, a signal must be seen in two detectors
with different ratios of $Z$ and $A-Z$.  However, to boost the
scattering cross-section and, therefore, neutralino detection
likelihood, many current experiments use heavy elements such as Ge
and Xe that have similar ratios $A-Z \approx 1.4 Z$.  It will be
necessary to see a signal in experiments using lighter elements
nearer to $A-Z \approx Z$ (\ie\ $Z \lae 17$) in order to determine
$\sigmanSI / \sigmapSI$.  Such experiments are possible, but their SI
sensitivities typically lag far behind those with heavier elements.

\begin{table}
  \begin{ruledtabular}
  \begin{tabular}{lccc}
    Model & C & L & M    \\
    \hline
    $\Deltaps$ = -0.15: & & & \\
    \hspace{2em}$\sigmapSD$ (pb) & 2.80 $\times 10^{-6}$
                                 & 2.25 $\times 10^{-6}$
                                 & 2.86 $\times 10^{-8}$ \\
    \hspace{2em}$\sigmanSD$ (pb) & 2.19 $\times 10^{-6}$
                                 & 1.69 $\times 10^{-6}$
                                 & 2.03 $\times 10^{-8}$ \\
    \hspace{2em}$\sigmanSD / \sigmapSD$ & 0.78
                                        & 0.75
                                        & 0.71 \\
    $\Deltaps$ = -0.12: & & & \\
    \hspace{2em}$\sigmapSD$ (pb) & 2.48 $\times 10^{-6}$
                                 & 2.03 $\times 10^{-6}$
                                 & 2.63 $\times 10^{-8}$ \\
    \hspace{2em}$\sigmanSD$ (pb) & 2.49 $\times 10^{-6}$
                                 & 1.89 $\times 10^{-6}$
                                 & 2.23 $\times 10^{-8}$ \\
    \hspace{2em}$\sigmanSD / \sigmapSD$ & 1.00
                                        & 0.93
                                        & 0.85 \\
    $\Deltaps$ = -0.09: & & & \\
    \hspace{2em}$\sigmapSD$ (pb) & 2.19 $\times 10^{-6}$
                                 & 1.82 $\times 10^{-6}$
                                 & 2.40 $\times 10^{-8}$ \\
    \hspace{2em}$\sigmanSD$ (pb) & 2.81 $\times 10^{-6}$
                                 & 2.10 $\times 10^{-6}$
                                 & 2.45 $\times 10^{-8}$ \\
    \hspace{2em}$\sigmanSD / \sigmapSD$ & 1.28
                                        & 1.15
                                        & 1.02 \\
    $\Deltaps$ = -0.06: & & & \\
    \hspace{2em}$\sigmapSD$ (pb) & 1.91 $\times 10^{-6}$
                                 & 1.63 $\times 10^{-6}$
                                 & 2.19 $\times 10^{-8}$ \\
    \hspace{2em}$\sigmanSD$ (pb) & 3.14 $\times 10^{-6}$
                                 & 2.33 $\times 10^{-6}$
                                 & 2.67 $\times 10^{-8}$ \\
    \hspace{2em}$\sigmanSD / \sigmapSD$ & 1.64
                                        & 1.43
                                        & 1.22 \\
    $\Deltaps$ = -0.03: & & & \\
    \hspace{2em}$\sigmapSD$ (pb) & 1.65 $\times 10^{-6}$
                                 & 1.44 $\times 10^{-6}$
                                 & 1.99 $\times 10^{-8}$ \\
    \hspace{2em}$\sigmanSD$ (pb) & 3.49 $\times 10^{-6}$
                                 & 2.57 $\times 10^{-6}$
                                 & 2.90 $\times 10^{-8}$ \\
    \hspace{2em}$\sigmanSD / \sigmapSD$ & 2.11
                                        & 1.78
                                        & 1.46 \\
    $\Deltaps$ = 0.00: & & & \\
    \hspace{2em}$\sigmapSD$ (pb) & 1.41 $\times 10^{-6}$
                                 & 1.26 $\times 10^{-6}$
                                 & 1.80 $\times 10^{-8}$ \\
    \hspace{2em}$\sigmanSD$ (pb) & 3.86 $\times 10^{-6}$
                                 & 2.81 $\times 10^{-6}$
                                 & 3.15 $\times 10^{-8}$ \\
    \hspace{2em}$\sigmanSD / \sigmapSD$ & 2.74
                                        & 2.23
                                        & 1.75 \\
  \end{tabular}
  \end{ruledtabular}
  \caption[$\Deltaps$]{
    The spin-dependent neutralino-nucleon scattering
    cross sections in the benchmark models for several values of
    $\Deltaps$.
    }
  \label{tab:Deltaps}
\end{table}

\subsection{\label{sec:SD} Spin-Dependent Parameters and
            Cross Sections}

The determination of the SD cross sections for a given MSSM model
depends on the three parameters specifying the spin content in a
nucleon: $\athree$, $\aeight$, and $\Deltaps$.  As demonstrated for
benchmark model C in \reftab{errors}, uncertainties in $\athree$
and $\aeight$ induce only $<1$\% and $\sim$ few \% uncertainties in
$\sigmaNSD$, respectively, at the 1-$\sigma$ level.  Uncertainties
in the strange spin contribution $\Deltaps$, on the other hand,
induce 10 to 15 \% uncertainties in $\sigmaNSD$, and uncertainties in
this parameter account for essentially all the width of the SD
confidence intervals in \reftab{models}.  Since uncertainties induced
by $\athree$ and $\aeight$ are negligible, we ignore these
terms and focus on $\Deltaps$.

We give in \reftab{Deltaps} the SD cross sections and their ratios
for the benchmark models for several $\Deltaps$ values ranging from
-0.15 (the 2-$\sigma$ lower bound) to 0.0 (no strange contribution
to the nucleon spin).  Over the 2-$\sigma$ range $-0.09 \pm 0.06$,
both $\sigmapSD$ and $\sigmanSD$ vary by $\sim$40\% in each of the
three models, a significant variation but not so large as
that induced in the SI cross sections by $\SigmapiN$.  However,
unlike $\SigmapiN$ and the SI cross sections, the SD proton and
neutron cross-sections are {\it anti}-correlated with 
$\Deltaps$: as the value of $\Deltaps$ increases, $\sigmapSD$
decreases whilst $\sigmanSD$ increases.

\begin{figure*}
  \includegraphics[width=1.00\textwidth]{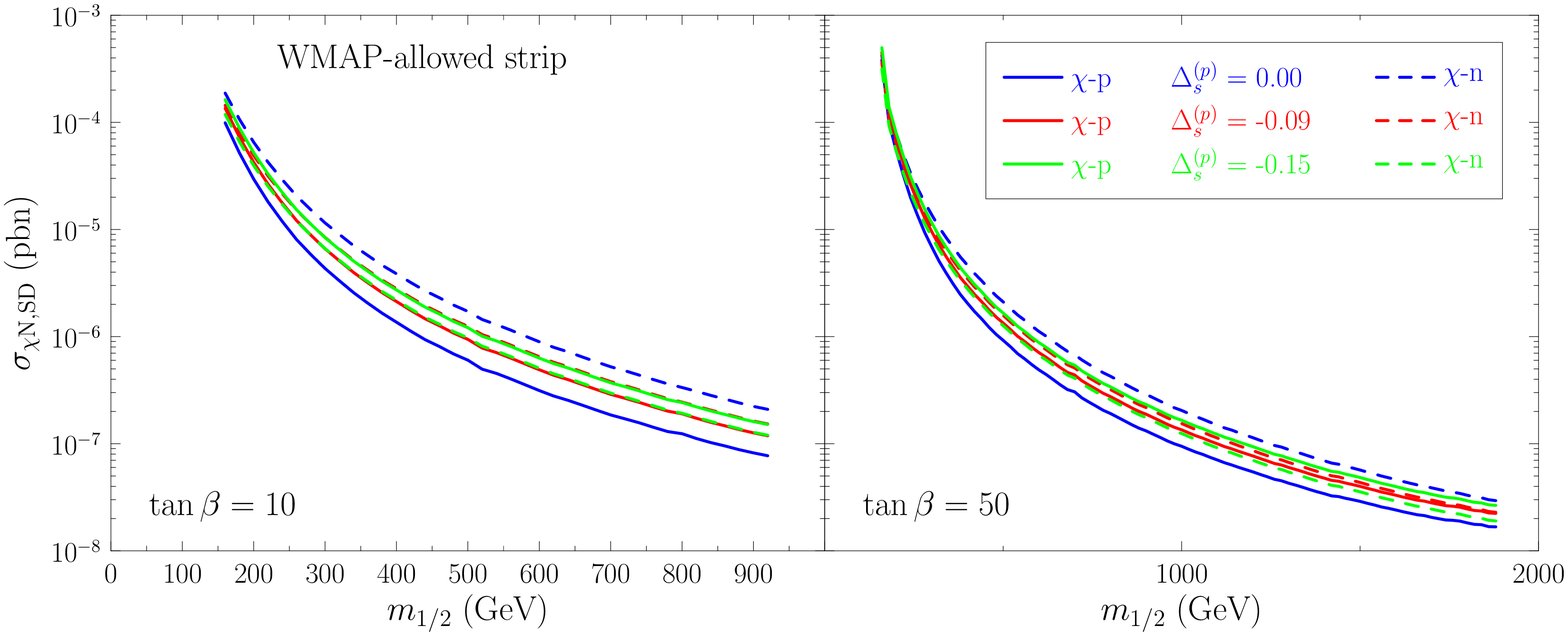}
  \caption[SD cross-section for several values of $\Deltaps$
           (WMAP strip)]{
    \coloronlinestatement{}
    The ratios of the spin-dependent neutralino-nucleon scattering
    cross sections along the WMAP-allowed strips for $\tanb = 10$
    (left panel) and $\tanb = 50$ (right panel) for several values of
    $\Deltaps$.
    }
  \label{fig:CSDeltaps}
\end{figure*}

\begin{figure*}
  \includegraphics[width=1.00\textwidth]{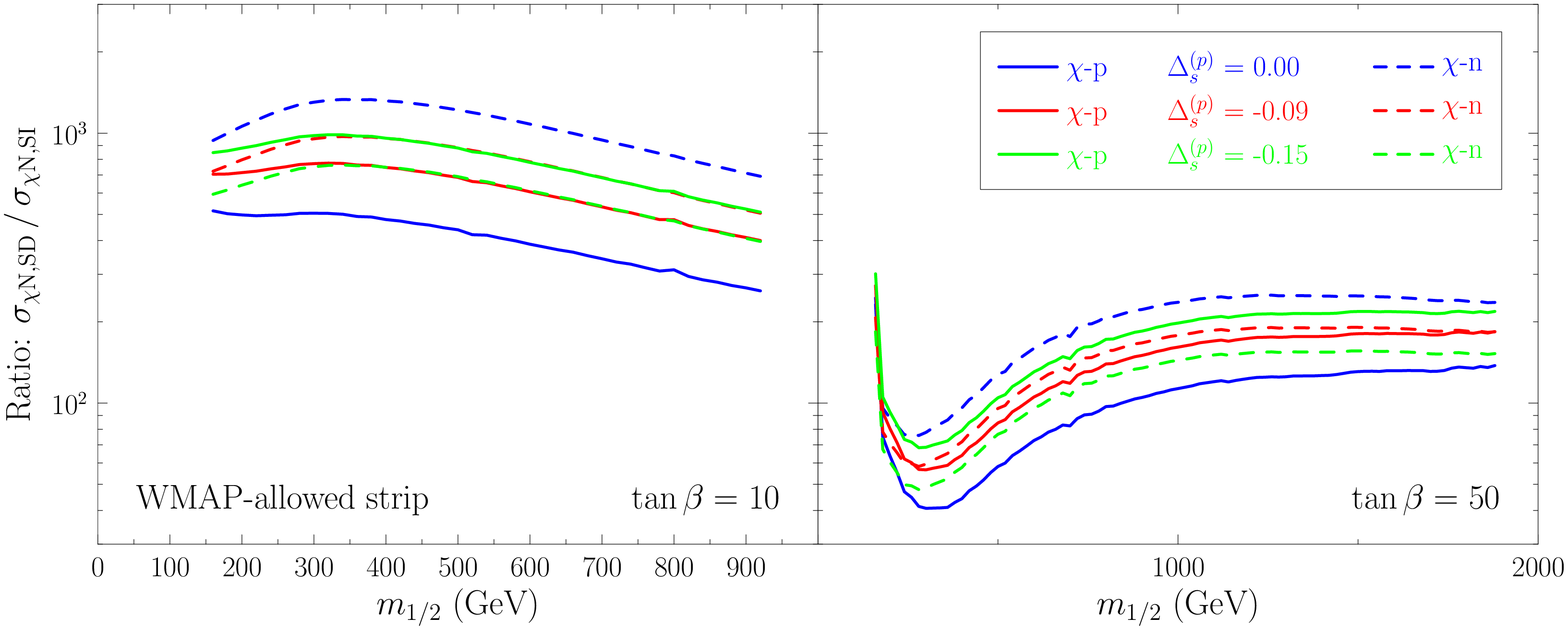}
  \caption[SD/SI cross-section ratio for several values of
           $\Deltaps$ (WMAP strip)]{
    \coloronlinestatement{}
    The ratios of the spin-dependent and spin-independent
    neutralino-nucleon scattering cross sections along the
    WMAP-allowed strips for $\tanb = 10$ (left panel) and
    $\tanb = 50$ (right panel) for several values of $\Deltaps$.
    }
  \label{fig:SDSIratioDeltaps}
\end{figure*}

\Reffig{CSDeltaps} and \reffig{SDSIratioDeltaps} show the SD
cross sections and the SD/SI cross-section ratios along the WMAP
coannihilation strips for several values of $\Deltaps$.
These figures again demonstrate the anti-correlation of the
neutron/proton cross sections with $\Deltaps$ as (for
decreasing $\Deltaps$) $\sigmapSD$ and $\sigmanSD$ approach each other
and, for low enough $\Deltaps$, cross over ($\sigmapSD$ becomes the
larger SD cross-section).

\begin{figure*}
  \includegraphics[width=1.00\textwidth]{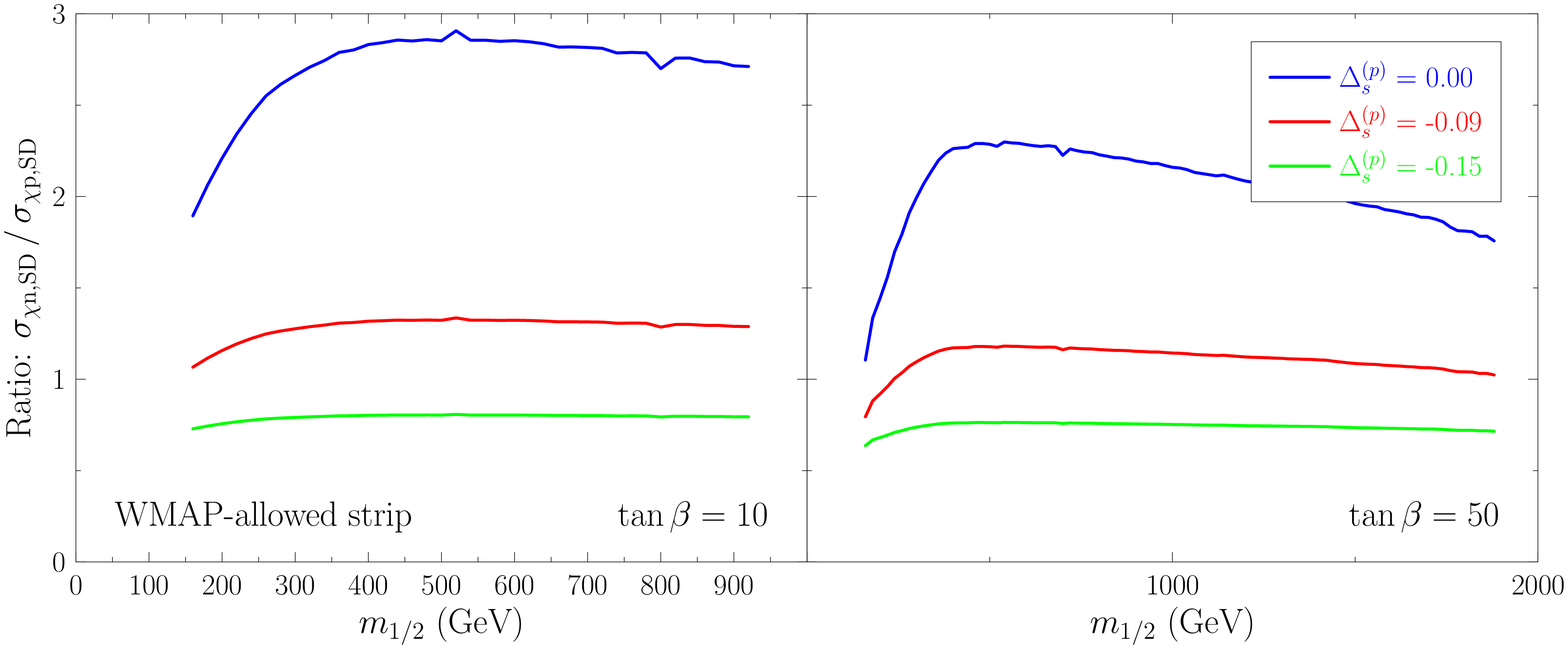}
  \caption[$\chi$-n/$\chi$-p cross-section ratio for several values of
           $\Deltaps$ (WMAP strip)]{
    \coloronlinestatement{}
    The ratios of the spin-dependent neutralino-neutron and
    neutralino-proton scattering cross sections along the
    WMAP-allowed strips for $\tanb = 10$ (left panel) and
    $\tanb = 50$ (right panel) for several values of $\Deltaps$.
    }
  \label{fig:npratioDeltaps}
\end{figure*}

Because of the anti-correlated behaviors of the SD cross sections, the
SD $\chi$-n/$\chi$-p cross-section ratio is particularly sensitive to
the value of $\Deltaps$.  This ratio is shown along the WMAP
coannihilation strips in \reffig{npratioDeltaps}, where it is apparent
that the ratio varies by a factor of 2 to 3 at the $\sim$~2-$\sigma$
level for $\Deltaps$.  That factor of 2 to 3 in the ratio can also be
seen in the confidence intervals at the 95.4 \% C.L., given in
\reftab{models}.

The large uncertainty in $\sigmanSD / \sigmapSD$ induced by
$\Deltaps$ for a given model is unfortunate since
this ratio may be one of the easiest to determine experimentally.
This is because many elements have a spin-odd proton group
($\langle S_{\rmpp} \rangle \not\approx 0$) and a spin-even
neutron group ($\langle S_{\rmpp} \rangle \approx 0$) or vice versa,
in which case $\Lambda \propto a_{\rmpp}$ or
$\Lambda \propto a_{\rmnn}$ in \refeqn{sigmaSD}.  By using an
odd-even element in a detector, any signal could
essentially be entirely attributed to $a_{\rmpp}$ and therefore yield
$\sigmapSD$.  Likewise, when using an even-odd element,
any signal could essentially be entirely attributed to $a_{\rmnn}$ and
therefore yield $\sigmanSD$.  Even though the measurements
would probably actually be of $\rhoDM \sigmapSD$ and $\rhoDM \sigmanSD$
and not of $\sigmapSD$ and $\sigmanSD$ alone, the ratio
$\sigmanSD / \sigmapSD$ is still an unambiguous and straightforward
experimental measurement.  The large uncertainties induced by
$\Deltaps$, however, make it difficult to use such a signal to
constrain the CMSSM parameter space.

\section{\label{sec:discussion} Summary and Discussion}

We have analyzed in this paper the principal hadronic uncertainties in
the spin-independent (SI) and spin-dependent (SD) cross sections for
supersymmetric relic scattering on protons and neutrons, using three
benchmark points and two coannihilation strips as illustrations. We
have found that the principal hadronic uncertainty in the SD cross
sections is due to our lack of knowledge of the $\pi$-nucleon $\sigma$
term. In comparison, uncertainties in quark masses and their ratios
are much less important. In the case of the SD cross sections, the
dominant uncertainty is due to our ignorance of the strange-quark
contribution to the nucleon spin, though this uncertainty is
relatively less important than that induced in the SI cross sections
by the $\pi$-nucleon $\sigma$ term.

This uncertainty in the $\pi$-nucleon $\sigma$ term clouds very
significantly the interpretation of searches for (and eventually
measurements of) dark matter scattering on nuclei, preventing
precise answers to the key questions: How do present
unsuccessful searches constrain the supersymmetric model parameter
space? How accurately could a possible future measurement be used to
refine the model parameters? This hadronic uncertainty is much larger
than that generates by uncertainties in supersymmetric model
calculations of the effective LSP-quark interactions, and also much
larger than the astrophysical uncertainty in the local cold dark
matter density.

One of the great hopes in supersymmetric phenomenology is that one
will eventually be able to use measurements at accelerators such as
the LHC and/or a linear $e^+ e^-$ collider to calculate the relic LSP
density in the Universe, and the rates for dark matter scattering.
The hadronic uncertainty in the latter that is induced by our
ignorance of the $\pi$-nucleon $\sigma$ term limits severely the
prospects for completing the second part of this programme.
Specifically, this uncertainty is much larger than the uncertainty in
calculating the relic LSP density that could be expected from LHC
measurements in at least one benchmark model.

{\it We therefore plead for an experimental campaign to determine
 better the $\pi$-nucleon $\sigma$ term.} This quantity is certainly
interesting and important in its own right and as a measure of the
importance of strange quarks in the nucleon. However, as argued in
this paper, it is potentially also a key ingredient in the effort to
understand one of the most important aspects of possible new physics
beyond the Standard Model.


\begin{acknowledgments}
  The work of KAO was supported in part by DOE Grant
  No.\ DE-FG02-94ER-40823.
  CS acknowledges the support of the William I.\ Fine Theoretical
  Physics Institute at the University of Minnesota
  and thanks L.~Duong for useful conversations.
\end{acknowledgments}




\end{document}